\documentclass[journal]{IEEEtran}

\usepackage[compress]{cite}
\usepackage[ruled, linesnumbered, titlenotnumbered, noend]{algorithm2e}
\usepackage{times,amssymb}
\usepackage{amsmath}
\usepackage{bbm}
\usepackage{subfigure}
\usepackage{bm}
\usepackage[font=small]{caption}
\usepackage[dvips]{graphicx}
\usepackage{url}
\usepackage{multirow}
\usepackage{booktabs}

\begin{document}
\title{Cost-efficient and Skew-aware Data Scheduling for Incremental Learning in 5G Networks}
\author{Lingjun~Pu,
        Xinjing~Yuan,
        Xiaohang~Xu,
        Xu~Chen,~\IEEEmembership{Senior Member,~IEEE,}
        Pan~Zhou,~\IEEEmembership{Senior Member,~IEEE,} and
        Jingdong~Xu.
\thanks{Manuscript received February 28, 2021; revised July 25, 2021; accepted August 28, 2021. This work was supported in part by the National Natural Science Foundation of China (No.62172241, No. U20A20159, No.61972448); the Technology Research and Development Program of Tianjin (No.18ZXZNGX00200); the Young Elite Scientists Sponsorship Program by Tianjin (No. TJSQNTJ-2018-19); the Open Project Fund of State Key Laboratory of Integrated
Services Networks (No. ISN2013); the Natural Science Foundation of Tianjin (20JCZDJC00610); the Program for Guangdong Introducing Innovative and Entrepreneurial Teams (No.2017ZT07X355). \emph{(Corresponding author: Xu Chen.)}
}
\thanks{
Lingjun Pu is with the College of Computer Science, Nankai University, Tianjin 300071, China, and also with
the State Key Laboratory of Integrated Services Networks, Xidian University, Xi'an 710126, China. (e-mail: pulingjun@nankai.edu.cn).}
\thanks{
Xinjing Yuan and Xiaohang Xu are with the College of Computer Science, Nankai University, Tianjin 300071, China.
(e-mail: \{yuanxj, xuxh\}@nankai.edu.cn).}
\thanks{
Xu Chen is with the School of Computer Science and Engineering, Sun Yat-sen University, Guangzhou 510006, China, and also with the Pazhou Lab, Guangzhou 510335, China.
(e-mail: chenxu35@mail.sysu.edu.cn).}
\thanks{
Pan Zhou is with the Hubei Engineering Research Center on Big Data Security and the School of Cyber Science and Engineering, Huazhong University of Science and Technology, Wuhan, 430074, China. (e-mail: panzhou@hust.edu.cn).}
\thanks{
Jingdong Xu is with the College of Computer Science, Nankai University, Tianjin 300071, China.
(e-mail: xujd@nankai.edu.cn).}
}
\maketitle
\begin{abstract}
To facilitate the emerging applications in 5G networks, mobile network operators will provide many network functions in terms of control and prediction. Recently, they have recognized the power of machine learning (ML) and started to explore its potential to facilitate those network functions. Nevertheless, the current ML models for network functions are often derived in an offline manner, which is inefficient due to the excessive overhead for transmitting a huge volume of dataset to remote ML training clouds and failing to provide the incremental learning capability for the continuous model updating. As an alternative solution, we propose \emph{Cocktail}, an incremental learning framework within a reference 5G network architecture. To achieve cost efficiency while increasing trained model accuracy, an efficient online data scheduling policy is essential. To this end, we formulate an online data scheduling problem to optimize the framework cost while alleviating the data skew issue caused by the capacity heterogeneity of training workers from the long-term perspective. We exploit the stochastic gradient descent to devise an online asymptotically optimal algorithm, including two optimal policies based on novel graph constructions for skew-aware data collection and data training. Small-scale testbed and large-scale simulations validate the superior performance of our proposed framework.
\end{abstract}
\begin{IEEEkeywords}
Incremental Learning, Data Scheduling, Skew Awareness, Cost Efficiency, 5G Networks.
\end{IEEEkeywords}

\section{Introduction}
In the coming 5G era, various kinds of attractive applications such as mobile AR/VR and autonomous driving are emerging and attracting great attention. To facilitate these novel applications, mobile network operators will not only deploy a massive number of base stations, pursue advanced cellular architectures, but also provide many network functions in terms of control and prediction \cite{agiwal2016next}.
In recent years, they have recognized the power of machine learning (ML) \cite{letaief2019roadmap, wang2020artificial} and started to explore its potential to facilitate network control functions (e.g., user association \cite{zappone2018user} and power allocation in Massive MIMO \cite{sun2018learning}) and network prediction functions (e.g., user CSI prediction \cite{luo2018channel} and cellular traffic prediction \cite{bega2019deepcog}). In order to apply ML to a specific network function, the most common method in the previous studies is to train an ML model with an offline dataset (i.e., ML training) in a centralized node (e.g., ML training cloud) and exploit the derived model to make online decision (i.e., ML inference). However, offline ML training suffers from the increasing number of base stations and mobile users, since there will be an excessive overhead and latency for transmitting a huge volume of dataset from cellular networks to remote ML training clouds. Moreover, it fails to provide the incremental learning capability for the continuous model updating with fresh data samples from the dynamic cellular network environments. Therefore, it is highly desirable to pursue an efficient incremental learning framework within the 5G network architecture.

\begin{figure*}[tt]
\centering
\includegraphics[height=2.8in]{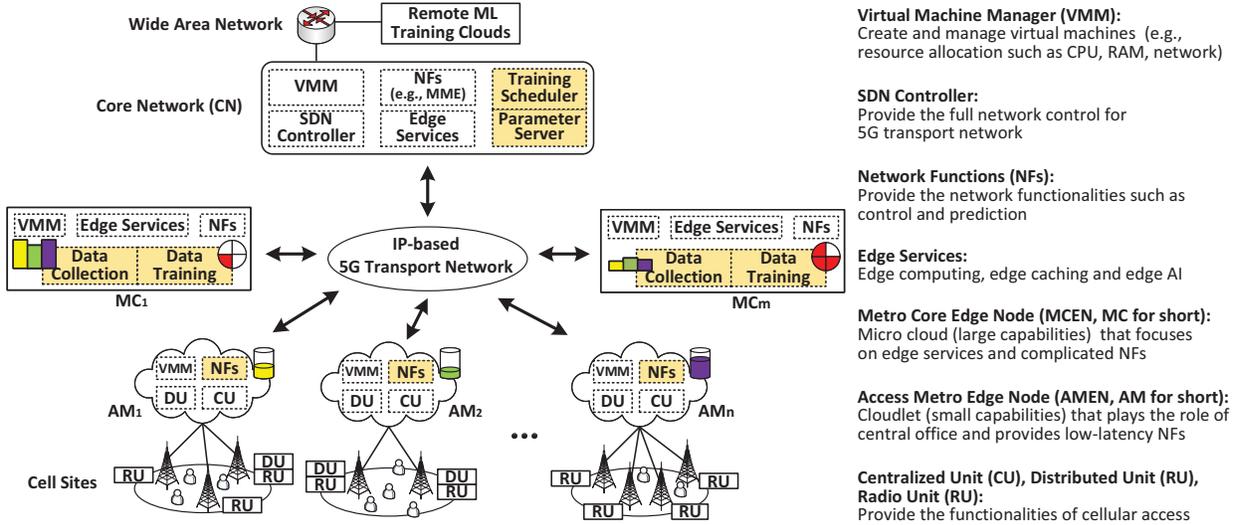}
\caption{The overview of the \emph{Cocktail} framework (the main components are marked with the yellow background).}
\label{fig:scenario}
\end{figure*}

Edge computing is an emerging and promising computing paradigm, where a number of micro clouds are deployed at the network edge to provide low-latency computing and caching services for mobile users, and it has become one of the most critical components of 5G networks \cite{taleb2017multi}. Inspired by the idea of edge computing, we propose \emph{Cocktail}, an incremental learning framework within a reference 5G network architecture (e.g., Metro-Haul \cite{MetroHaul} and Cloud-native 5G \cite{Cloud5G}). As shown in Fig. \ref{fig:scenario}, we consider the parameter server based ML framework \cite{li2014scaling}, in which a set of access metro edge nodes (AMs: data sources\footnote{We consider AM as data source, since it will provide low-latency network functions for base stations and mobile users. Applying ML to these functions has become the main trend in the 5G era \cite{letaief2019roadmap, wang2020artificial}.}) are selected to continuously generate data samples for a given network function offered by themselves; a set of metro core edge nodes (MCs: training workers) are selected to cooperatively train an evolved ML model with the collected data samples from the selected AMs in an online manner; in the core network, the parameter server is responsible for aggregating local ML model from each training worker, and the training scheduler is responsible for continuously making online decision for each training worker on how many data samples it will collect from the data sources (i.e., data collection) and how many collected data samples from each data source it will train (i.e., data training).

Despite the great potential, designing an efficient data scheduling (i.e., data collection and data training) policy for the training scheduler in our framework remains challenging:

(1) \emph{It is not a one-shot operation but required to continuously adapt to the system dynamics}. The main reason is that the network capacity between each pair of AM and MC as well as the computing capacity of each MC allocated to our framework is limited and time-varying. To decouple the mutual restriction between data collection and data training, we will introduce a set of AM queues in each MC to maintain the collected data samples from different AMs. In the context of queueing, the decision making in the training scheduler is time-correlated.

(2) \emph{It should involve a mechanism to alleviate the data skew issue \cite{zhao2018federated, krawczyk2016learning} caused by the capacity heterogeneity of training workers (i.e., MCs)}. From the perspective of an individual MC, if the network capacity between it and a specific AM achieves a higher value over a period of time, the backlog of its corresponding AM queue will accumulate quickly. To fully utilize the computing capacity, this MC would train more data samples from that AM queue over time, leading to a skewed data training, which could adversely impact the accuracy of local ML model. From the perspective of overall MCs, if some MCs have a low computing capacity over a period of time, the number of trained data samples from their AM queues is far less than the other MCs', leading to a low-effective parameter aggregation, which could prolong the convergence time of global ML model.

(3) \emph{It should make our framework cost-efficient and scalable}. Intuitively, the proposed incremental learning framework should not introduce great overhead in terms of transmission and computation, which is also in accordance with many edge services such as computing and caching \cite{pu2018online, mao2017survey}. Besides, it should be scalable to accommodate a large number of AMs (i.e., data sources) and MCs (i.e., training workers) for the diversity of 5G networks.

To address the above challenges, we first build a comprehensive model to capture the specific behavior of data sources (AMs), training workers (MCs), parameter server and training scheduler in our framework. To tackle the data skew issue, we advocate the ML training with
worker cooperation (i.e., data offloading), and also introduce a long-term skew amendment mechanism. Then, we formulate an online data scheduling problem to optimize the framework cost while reconciling the data skew issue from the long-term perspective (Section III).

We exploit the stochastic gradient descent technique to devise an online data scheduling algorithm, including two optimal policies based on novel graph constructions for skew-aware data collection and skew-aware data training per time slot. Theoretical analysis shows that the proposed online algorithm achieves an asymptotic optimum with polynomial running time. In addition, we improve the proposed online algorithm with online learning to speedup the convergence of in-network training (Section IV).

We implement a small-scale testbed and adopt a cellular traffic prediction task (i.e., prediction function) by using realistic dataset to evaluate the performance of our proposed algorithm. The evaluation results are in accordance with the theoretical analysis, and they also show that our proposed algorithm can achieve up to 1.7$\times$ convergence time reduction by taking long-term skew amendment mechanism and online learning mechanism into account. In addition, we also
conduct large-scale simulations and adopt a base station power allocation task (i.e., control function) by using synthetic dataset to evaluate the performance. The evaluation results show that our proposed algorithm can achieve up to 78.5\% framework cost reduction compared with alternative solutions (Section V).

\section{Related Work}
\textbf{Centralized Machine Learning (CML) \cite{chu2007map}}. In the CML framework, a centralized node (e.g., cloud) will periodically collect data samples from different data sources and train the ML model in a map-reduce manner. That is, the centralized node evenly partitions the collected dataset into multiple subsets, distributes each subset to a resource-sufficient server to train a local ML model and aggregates the parameters of all the local ML models as the global ML model. The similarity between \emph{Cocktail} and CML is that the centralized node (i.e., the training scheduler) has the global information about data sources (i.e., AMs) and training workers (i.e., MCs), while their key difference is that the training workers in \emph{Cocktail} directly collect data from data sources rather than receiving the distributed ones from the centralized node. Note that the CML framework is inefficient for in-network training, since data collection in CN and then distribution to multiple MCs will lead to great transmission overhead.

\textbf{Distributed Machine Learning (DML) \cite{10.1145/3377454}}. In the DML framework, each distributed training worker will periodically collect data samples from its nearby data sources, train a local ML model and synchronize it with the parameter server \cite{li2014scaling, hsieh2017gaia} or by workers' exchange \cite{watcharapichat2016ako, hong2020dlion}. The similarity between \emph{Cocktail} and DML is that the training workers directly collect data samples from data sources, while their key difference is that the training workers in \emph{Cocktail} can collect the data samples from all data sources rather than their nearby ones. In addition, many DML works assume the data collected by each training worker follows an independent identically distribution (IID), and also assume the computing capacity of each training worker is sufficient. However, these assumptions no longer hold in our framework (e.g., the computing capacity of MCs allocated to our framework is limited).

\textbf{Federated Learning (FL) \cite{lim2019federated}}. The FL framework is a special DML framework, which allows an ML model to be trained on end devices in a distributed manner. Unlike CML and DML, it is faced with several challenges such as the limited resources of end devices and the non-IID of local dataset generated by each device. As a common idea, only a subset of devices is selected to participate in each training round. For example, FedAvg \cite{mcmahan2017communication} randomly selects a subset of devices in each round and aggregates their local ML models to update the global ML model. The authors in \cite{xia2020multi, wang2020optimizing} respectively exploit multi-arm bandit and reinforcement learning to select the proper training workers, so as to counterbalance the biases introduced by the non-IID data from end devices. Although \emph{Cocktail} is faced with the similar challenges, the main differences between \emph{Cocktail} and FL are two-fold. First, each worker in FL is also the data source, which could inevitably generate the non-IID data samples, and the common idea to relieve the non-IID issue is to strategically select a subset of workers to participate in each training round. However, each worker in \emph{Cocktail} will collect the data samples from multiple data sources, and a wise data scheduling scheme is proposed to avoid the skewed data training. In other words, their strategies to tackle the skewed data training are completely different (i.e., no data collection + worker selection vs. data scheduling). Second, the workers in FL do not cooperate due to privacy issue, while our framework advocates the ML training with worker cooperation.

\textbf{Machine Learning with Edge Computing}. There has been a growing interest in marrying ML with edge computing during the last five years (see the surveys \cite{zhou2019edge, wang2020convergence} as references). However, most of works focus on the edge-assisted model inference. Recently, some researchers start to consider FL in edge networks (see the survey \cite{lim2019federated} as a reference), and they mainly explore the network and computing resource allocation to benefit local training \cite{tran2019federated, abad2020hierarchical}, the incentive mechanism to recruit the participants \cite{nishio2019client, kang2020reliable} and the global aggregation to accelerate the training convergence under a given resource budget \cite{wang2018edge}. Although these works also consider the centralized control, the specific setting (e.g., worker cooperation) and key problem (i.e., data collection and data training) considered in ours are completely different from them.

\section{Framework Model and Problem Formulation}
\quad As shown in Fig. \ref{fig:scenario}, the \emph{Cocktail} framework is built within a reference 5G network architecture (e.g., Metro-Haul \cite{MetroHaul} and Cloud-native 5G \cite{Cloud5G}), where Access Metro Edge Node (AM) plays the role of traditional central office and provides many latency-sensitive network functions for base stations and mobile users such as cellular traffic prediction and user association; Metro Core Edge Node (MC) provides fruitful edge services (e.g., edge computing and caching) and some complicated network functions (e.g., deep packet inspection); Core Network (CN) provides the full network control and orchestration with the SDN controller in it and the virtual machine manager (VMM) in the above three entities (i.e., AM, MC and CN).
In our framework, we consider AM as data source, since it provides low-latency network functions for base stations and mobile users \cite{letaief2019roadmap, wang2020artificial}. We consider MC as training worker, since it has much larger computing and network capabilities compared with AM. We consider CN as centralized controller, since it can obtain the global information of all AMs and MCs. {In practice, AMs, MCs and CN are connected by an IP-based 5G transport network \cite{MetroHaul,Cloud5G}. In this context, they can exploit the socket-based communication to exchange information. For example, an AM will adopt unicast to upload data samples to a MC, and the CN will adopt multicast to transmit the global model to the selected AMs in each training round.
\begin{figure}[thb]
\centering
\includegraphics[height=1.8in]{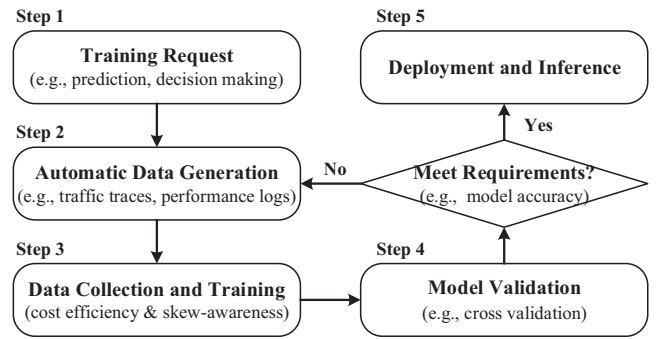}
\caption{The main procedure of the \emph{Cocktail} framework.}
\label{fig:flow}
\end{figure}

The basic procedure of our framework is shown in Fig. \ref{fig:flow}. When CN receives a model training request tailored for a given network function offered by AM, it will create a virtual machine (VM) to act as both the training scheduler and the parameter server. Then, the training scheduler will select a set $\mathcal{N}=\{1,2,\dots,N\}$ of AMs to generate the required data samples\footnote{In this paper, we use data and data sample interchangeably.} in terms of the given network function\footnote{In practice, the data samples are automatically generated by the given network function itself in each selected AM, and we will further discuss it in Section III-A.}. Next, it will select a set $\mathcal{M}=\{1,2,\dots,M\}$ of MCs\footnote{The VMM of each selected MC will create a VM for our framework.} to continuously collect data samples from the selected AMs (i.e., data collection), update local ML model (i.e., data training) and synchronize it with the parameter server. We consider such an in-network training operates in a time-slotted pattern (e.g., at the timescale of tens of minutes). In other words, the training scheduler will periodically make decisions on data collection and data training for each training worker. For simplicity, we assume each slot $t$ has a unit length. When the global ML model is converged (e.g., the objective of the loss function is lower than a given threshold or the total training round is larger than a given threshold), the parameter server will conduct sample testing to evaluate the accuracy of the global ML model. If the model accuracy is acceptable, it will inform the selected AMs (MCs) to terminate data generation (data collection and data training), and then deploy the global ML model to all the AMs\footnote{We should emphasize that the AMs and MCs are established and managed by a cellular network operator, and therefore they are trustworthy in general. In addition, if the operation behaviors of AMs for a given network function are different (i.e., they require different ML models), we could first classify the AMs with the similar operating behavior into several AM clusters, and then select some AMs belonging to the same AM cluster as data sources for the required model training.} as the decision maker for the corresponding network function. If the model accuracy is unacceptable, the parameter server will restart the model training by selecting new AMs and MCs while taking the trained global ML model as the initial model in each selected MC. To facilitate the following reading, we summarize the main notations in Table I.

\begin{table}[tt]  
\centering  
\label{tab1}
 \begin{tabular}{ c l }  
     \toprule
     {Parameters} & {Descriptions}  \\
     \midrule
     {$\mathcal{N}$} & {the set of selected AMs (i.e., data sources)}\\
     \specialrule{0em}{1pt}{1pt}
     {$\mathcal{M}$} & {the set of selected MCs (i.e., training workers)}\\
     \specialrule{0em}{1pt}{1pt}
     {$Q_i(t)$} & {the data queue of AM $i$ storing its generated data}\\
     \specialrule{0em}{1pt}{1pt}
     {$\zeta$} & {the expected data generation rate of each selected AM}\\
     \specialrule{0em}{1pt}{1pt}
     {$A_i(t)$} & {the instantaneous number of generated data of AM $i$}\\
     \specialrule{0em}{1pt}{1pt}
     {$d_{ij}(t)$} & {the network capacity between AM $i$ and MC $j$}\\
     \specialrule{0em}{1pt}{1pt}
     {$R_{ij}(t)$} & {the AM queue in MC $j$ that stores the collected data} \\
     & {from AM $i$}\\
     \specialrule{0em}{1pt}{1pt}
     {$f_j(t)$} & {the computing capacity of MC $j$}\\
     \specialrule{0em}{1pt}{1pt}
     {$D_{jk}(t)$} & {the network capacity between MC $j$ and MC $k$}\\
     \specialrule{0em}{1pt}{1pt}
     {$\rho$} & {the computing resource consumption}\\
     & {of training one data sample}\\
     \specialrule{0em}{1pt}{1pt}
     {$\Omega_{ij}(t)$} & {the number of data from AM $i$ trained by MC $j$}\\
     \specialrule{0em}{1pt}{1pt}
     {$\mathcal{C}(t)$} & {the framework cost including} \\
     & {data collection cost between AM and MC},\\
     & {data offloading cost between MCs},\\
     & {and data training cost in MCs}\\
     \midrule
     {Variables} & {Descriptions}  \\
     \midrule
     {${\alpha_{ij}(t)\!\in\!\{0,1\}}$} & {the control variable indicating whether AM $i$ will} \\
     & {establish a connection to MC $j$}\\
     \specialrule{0em}{1pt}{1pt}
     {${\theta_{ij}(t)\!\in\![0,1]}$} & {the control variable indicating the connection duration} \\
     & {between AM $i$ and MC $j$}\\
     \specialrule{0em}{1pt}{1pt}
     {${x_{ij}(t)\in\mathbb{R}}$} & {the control variable indicating the number of data}\\
     & {from AM queue $R_{ij}(t)$ in MC $j$ trained by MC $j$}\\
     \specialrule{0em}{1pt}{1pt}
     {${y_{ijk}(t)\in\mathbb{R}}$} & {the control variable indicating the number of data} \\
     & {from AM queue $R_{ij}(t)$ in MC $j$ offloaded to MC $k$}\\
     \specialrule{0em}{1pt}{1pt}
     {${z_{jk}(t)\!\in\!\{0,1\}}$} & {the control variable indicating whether MC $j$ will}\\
     & {establish a connection to MC $k$}\\
     \specialrule{0em}{1pt}{1pt}
     \bottomrule
 \end{tabular}
 \caption{{Main notations in the \emph{Cocktail} framework}.}  
\end{table}

\subsection{AM (Data Source) Model}\label{AM}
\textbf{Data Generation}. We consider the supervised learning: a data sample consists of an input vector for the ML model and a desired scalar output. In our framework, data samples are automatically generated by the selected AMs without human supervision.
Specifically, concerning a network control function, its data sample is derived by optimally solving the corresponding optimization problem. Taking base station power allocation \cite{sun2018learning} as an example, the input of a data sample could be the signal-to-noise ratio (SNR) between each user and the base station. The output could be the optimal power allocation for each user in terms of some metrics\footnote{
Such a data generation is reasonable in practice, since many network control functions nowadays adopt algorithmic strategies and that the data generation could be in parallel with the regular network control (e.g., a network function exploits a heuristic algorithm to make realtime control with a low latency while using the optimal algorithm to generate data samples).} (e.g., the sum-rate maximization). Concerning a network prediction function, its data sample is easily derived in terms of network logs. Taking cellular traffic prediction \cite{bega2019deepcog} as an example, the network function can record a time series of cellular traffic through AM. If the ML model adopts recurrent neural networks (e.g., LSTM), then the input of a data sample could be a number of consecutive traffic records and the output could be the traffic record thereafter.

For a given model training request, we consider each data sample has an identical size\footnote{Unless otherwise noted, the unit of each variable in this paper is the size of one data sample.} (e.g., tens of KBs) and each AM $i\in\mathcal{N}$ maintains a data queue $Q_i(t)$ to store the generated data samples. We denote by $\zeta$ the expected data generation rate of each selected AM, which is set by the parameter server in CN and by $A_i(t)$ the instantaneous number of generated data of AM $i$ (i.e., $\mathbb{E}[A_i(t)]\!=\!\zeta$). We assume each AM $i$ has accumulated a sufficient number of data samples before the in-network training operates. That is, $Q_i(0)=Q_0, \forall i\!\in\!\mathcal{N}$. Note that our framework will consider the data samples from all the selected AMs due to their different temporal and spatial features for the global ML model training.

\subsection{MC (Training Worker) Model}\label{MC}
\textbf{Data Collection}. We consider that each MC can communicate with all the selected AMs, since they are connected by the IP-based 5G transport network \cite{MetroHaul,Cloud5G}, and it will collect and train the data samples from all of them due to their different temporal and spatial features. For ease of implementation and management, we consider each MC will exploit time division multiplexing to communicate with the selected AMs, and each AM can establish and maintain at most one connection to MCs, due to the limited AM bandwidth and its competition among different cellular network services \cite{mao2017survey}. To this end, we denote by $d_{ij}(t)$ the network capacity between AM $i$ and MC $j$, by $\alpha_{ij}(t)$ the control variable indicating whether AM $i$ will establish a connection to MC $j$, and by $\theta_{ij}(t)$ the control variable indicating the connection duration in slot $t$. Then, we have the following queue dynamics of $Q_i(t)$ and MC network capacity constraint:
\begin{align}
& \mathsmaller{Q_i(t\!+\!1)=\Big[Q_i(t)-\sum\nolimits_{j}\alpha_{ij}(t)\theta_{ij}(t)d_{ij}(t)\Big]^+\!+\!A_i(t)},\\
& \qquad \qquad \qquad\mathsmaller{\sum\nolimits_{j\in\mathcal{M}}\alpha_{ij}(t)\le1},\\
& \qquad \qquad \qquad\mathsmaller{\sum\nolimits_{i\in\mathcal{N}}\theta_{ij}(t)\le1},
\end{align}
where $[x]^+\triangleq\max\{x,0\}$ and $\theta_{ij}(t)d_{ij}(t)$ indicates the number of data samples uploaded from AM $i$ to MC $j$ in slot $t$. Constraint (2) indicates that each AM can establish at most one AM\! --\! MC connection. Constraint (3) indicates that the total duration of AM connections of one MC cannot exceed the slot length. In order to fully utilize MC computing and network capacity, we consider each MC $j$ maintains a data queue $R_{ij}(t)$ for each AM $i$ to store the collected data samples. We call \{$R_{ij}(t), \forall i\!\in\!\mathcal{N}$\} by AM queues in MC $j$. Initially, $R_{ij}(0)=0, \forall i, \forall j$. We should emphasize that AM $i$ in practice will exploit unicast to upload a number of data samples to MC $j$ in terms of the result of $\alpha_{ij}(t)\theta_{ij}(t)d_{ij}(t)$ in slot $t$. Besides, our framework and the proposed data scheduling algorithm can easily adapt to different connection settings, such as each AM can establish and maintain concurrent connections to multiple MCs (i.e., removing the binary control variable $\alpha_{ij}(t)$).

\textbf{Local Training}. As for the per-slot local training, we denote by $\mathcal{D}_j(t)$ the set of data samples trained by MC $j$, by $\mathcal{W}_j(t)$ the derived parameters (weights) of local ML model, and by $F_j(\mathcal{W}_j(t))$ the local loss function in slot $t$. According to the existing distributed machine learning frameworks \cite{wang2018edge, hsieh2017gaia}, each MC initially has the same parameters (i.e., $\mathcal{W}_j(0)=\mathcal{W}_0, \forall j\!\in\!\mathcal{M}$) and then exploits a gradient descent technique to update the parameters in terms of the trained dataset and the local loss function. Formally, we have
\begin{align} \label{C4}
{\mathcal{W}_j(t+1)=\mathcal{W}_j(t)-\tau\frac{\sum\nolimits_{h\in\mathcal{D}_j(t)}\nabla_{h}F_j(\mathcal{W}_j(t))}{|\mathcal{D}_j(t)|}},
\end{align}
where $|\mathcal{D}_j(t)|$ is the set size, $\nabla_{h}F_j(\mathcal{W}_j(t))$ is the gradient of $F_j(\mathcal{W}_j(t))$ based on a data sample $h\in\mathcal{D}_j(t)$, and $\tau$ is the step-size of the gradient descent.

We consider that a good trained dataset in our framework should consider the following issues: \emph{sample reliability}, \emph{sample diversity} and \emph{sample evenness}.
In our framework, since the data samples are automatically generated by traffic traces or performance logs, we believe they are of high reliability. In addition, each MC (i.e., training worker) can communicate with all the selected AMs (i.e., data sources), since they are connected by the IP-based 5G transport network \cite{MetroHaul,Cloud5G}, and it will collect and train the data samples from all of them due to their different temporal and spatial features. Therefore, we consider the collected dataset in each MC is of large diversity. Moreover, the network capacity between each pair of MC and AM is generally different, due to different hop counts and competitions among cellular network services, resulting in different numbers of collected data samples from AMs in each MC (i.e., non-identical sample distribution from data sources).
In this context, we will mainly consider the sample evenness issue. That is, each MC should evenly train the data from its AM queues as far as possible, since the trained model will eventually be used for all the AMs in 5G networks.} To achieve a good sample evenness (i.e., model generalization), we advocate a cooperative local training. In other words, if an MC lacks data samples from some AMs, it can ``borrow" some data from the corresponding AM queues in another MC\footnote{Note that we do not suggest the AMs to replenish the MC during its local training, since the network capacity between MCs is much larger and has less competition among different cellular network services \cite{mao2017survey}.} (i.e., data offloading). Note that the MC which ``lends" some data samples to another MC will remove those data from its AM queues, in order to ensure that each data sample will be trained only one time in our framework.

To proceed, we denote by $f_j(t)$ the computing capacity of MC $j$ and by $D_{jk}(t)$ the network capacity between MC $j$ and MC $k$. They may change across slots, due to the resource competition between our framework and other edge services or network functions in MCs.
In addition, we introduce two control variables $x_{ij}(t), y_{ijk}(t)$ to respectively indicate the number of data from AM queue $R_{ij}(t)$ in MC $j$ trained by MC $j$ itself and offloaded to MC $k$ in slot $t$. Besides, we introduce the control variable $z_{jk}(t)$ to indicate whether MC $j$ will establish a connection to MC $k$, and assume each MC will establish and maintain at most one MC\! --\! MC connection in slot $t$. We consider such a one-connection setting is easily implemented and sufficient in a relatively long time slot (i.e., at the timescale of tens of minutes) in our framework. Then, we have the following MC\! --\! MC transmission constraints:
\begin{align} \label{C6}
& \qquad \mathsmaller{\sum\nolimits_{k\in\mathcal{M}}z_{jk}(t)+\sum\nolimits_{m\in\mathcal{M}\setminus k}z_{mj}(t)\le1},\\
&  \qquad \mathsmaller{\sum\nolimits_{i\in\mathcal{N}_j}y_{ijk}(t)+\sum\nolimits_{i'\in\mathcal{N}_k}y_{i'kj}(t)\le D_{jk}(t)},\\
&  \qquad \qquad\qquad \mathsmaller{y_{ijk}\le z_{jk}},
\end{align}
\noindent where constraint (5) indicates that each MC can establish at most one MC\! --\! MC connection, and it implicity requires $z_{jk}(t)\!=\!z_{kj}(t)$. Constraint (6) indicates that the total number of offloaded data between two MCs cannot exceed the network capacity between them.
Constraint (7) indicates that the data offloading between two MCs is valid only if they have connected to each other.
Moreover, let $\rho$ denote the computing resource consumption of training one data sample, then we derive the MC training constraints:
\begin{align} \label{C8}
\mathsmaller{\sum\nolimits_{i\in\mathcal{N}}\big[x_{ij}(t)+\sum\nolimits_{k\in\mathcal{M}\setminus j}y_{ikj}(t)\big]\rho\le f_j(t)},
\end{align}
\noindent where constraint (\ref{C8}) indicates that the total computing resources consumed for training the dataset $\mathcal{D}_j(t)$ cannot exceed the available computing capacity of MC $j$ per slot. Intuitively, $|\mathcal{D}_j(t)|=\sum\nolimits_{i}\big[x_{ij}(t)+\sum\nolimits_{k}y_{ikj}(t)\big]$.

\textbf{Long-term Skew Amendment}. Although we introduce the cooperative local training to facilitate the sample evenness, we have not provided a specific metric to measure it. In addition, as we consider the incremental training, the time-varying system state $S(t)=\{d_{ij}(t), D_{jk}(t), f_j(t), \forall i, \forall j,$ $\forall k\}$ could make the sample evenness even harder.
For example, if the network capacity between a MC and a specific AM is always higher, the backlog of the corresponding AM queue in the MC will accumulate quickly. To fully utilize the computing capacity, the MC would train more data samples from that AM queue over time, leading to a long-term skewed data training, which would adversely impact the accuracy of local ML model.
In addition, if some MCs have a low computing capacity over a period of time, the number of data samples trained by them is far less than the other
MCs', leading to a low-effective parameter aggregation in the parameter server, which could prolong the convergence time of global ML model.
To this end, we should tackle the data skew issue from a long-term perspective.
In general, \emph{a sufficiently large subset, in which the proportion of data from any data source approximates the data proportion in the whole dataset, can well represent the dataset for the ML training}. With this principle, we have the long-term data skew constraint for each pair of AM $i$ and MC $j$ as follows:
\begin{align} \label{C9}
\Big\arrowvert\frac{\sum\nolimits_t\Omega_{ij}(t)}{\sum\nolimits_t\sum\nolimits_{l\in\mathcal{N}}\Omega_{lj}(t)}-\frac{1}{N}\Big\arrowvert\le\delta,
\end{align}
where $|x|$ is the absolute value of $x$, $\Omega_{ij}(t)$ refers to the number of data from AM $i$ trained by MC $j$ in slot $t$, i.e., $\Omega_{ij}(t)=x_{ij}(t)+\sum\nolimits_k y_{ikj}(t)$,
$\sum\nolimits_{l\in\mathcal{N}}\Omega_{lj}(t)$ refers to the total number of data trained by MC $j$ in slot $t$,
$1/N$ can be approximately viewed as the data proportion of each AM in the whole dataset, and $\delta$ indicates the tolerance of data skew. In practice, we can set the value of $\delta$ to 1\%$-$10\%. Note that such a ``probability" distance for the data distribution on each MC compared
with the global distribution is also the main indicator of the divergence of trained model parameters \cite{zhao2018federated}. To facilitate the following discussions, we rewrite constraint (\ref{C9}) as follows:
\begin{align}\label{C10}
& \mathsmaller{\check{\delta}_i\overline{\sum\nolimits_l[x_{lj}(t)\!+\!\sum\nolimits_k y_{lkj}(t)]}\!\le\!\overline{x_{ij}(t)\!+\!\sum\nolimits_k y_{ikj}(t)}},\\
& \mathsmaller{\overline{x_{ij}(t)\!+\!\sum\nolimits_k y_{ikj}(t)}\!\le\!\hat{\delta}_i\overline{\sum\nolimits_l[x_{lj}(t)\!+\!\sum\nolimits_k y_{lkj}(t)]}},
\end{align}
where $\check{\delta}_i\triangleq\frac{1}{N}-\delta$, $\hat{\delta}_i\triangleq\frac{1}{N}+\delta$ and the overline refers to the time-average operation. We should emphasize that constraint (9) can also be viewed as a metric to measure the sample evenness.
That is, if constraint (9) holds for each pair of AM and MC, then we can conclude the trained dataset in each MC achieves a good long-term sample evenness.

In the end, we provide the queue dynamics of $R_{ij}(t)$ and its associated constraint as follows:
\begin{align} \label{C12}
&\mathsmaller{R_{ij}(t\!+\!1)\!=\!\Big[R_{ij}(t)\!-\!x_{ij}(t)\!-\!\sum\nolimits_{k}y_{ijk}(t)\Big]^+}\notag\\
&\qquad \qquad \qquad \qquad\qquad \qquad +\alpha_{ij}(t)\theta_{ij}(t)d_{ij}(t),\\
&\mathsmaller{x_{ij}(t)+\sum\nolimits_{k}y_{ijk}(t)\le R_{ij}(t)},
\end{align}
where constraint (13) indicates that the total number of data locally trained plus offloaded cannot exceed the current queue backlog\footnote{We do not consider such a constraint for the queue $Q_i(t), \forall i$, since AMs as data sources have accumulated a sufficient number of data samples (i.e., we implicitly have that $Q_i(t)\!\gg\!d_{ij}(t), \forall i, \forall j, \forall t$).}.
After the training scheduler makes decisions on $x_{ij}(t), y_{ijk}(t)$ and $z_{jk}(t), \forall i, \forall j, \forall k$ in slot $t$, each MC $j$ will generate the dataset $\mathcal{D}_j(t)$ and exploit equation (\ref{C4}) to update local ML model. Note that the instantaneous
values of $x_{ij}(t), y_{ijk}(t)$ are sufficiently large in a time slot (i.e., tens of minutes), and therefore we allow them to take real numbers as a simple approximation.

\subsection{CN (Centralized Controller) Model}\label{CN}
\textbf{Training Scheduler}. The training scheduler in our framework is responsible for estimating the system state $S(t)$ and maintaining the instantaneous queue state $Q_i(t), R_{ij}(t), \forall i, \forall j$ as well as the long-term data skew state $\Omega_{ij}(t), \forall i, \forall j$, to make decisions on data collection (i.e., $\alpha_{ij}(t), \theta_{ij}(t),$ $\forall i, \forall j$) and data training (i.e., $x_{ij}(t), y_{ijk}(t), z_{jk}(t), \forall i, \forall j, \forall k$) for all the MCs per slot. The main purpose of the centralized decision making is to optimize the framework cost while alleviating the data skew issue from the long-term perspective.

We mainly consider the data collection cost between each pair of AM and MC, data offloading cost between MCs, and data training cost in MCs, since they are directly affected by the decision making. In this context, we exploit $\mathcal{C}(t)$ to denote the overall framework cost which is given by:
\begin{align} \label{C14}
&\mathsmaller{\mathcal{C}(t)=\sum\nolimits_{i}\sum\nolimits_{j}c_{ij}(t)\alpha_{ij}(t)\theta_{ij}(t)d_{ij}(t)}\notag\\
&\mathsmaller{\qquad+\sum\nolimits_{j}\sum\nolimits_{k}e_{jk}(t)\sum\nolimits_{i}y_{ijk}(t)}\notag\\
&\mathsmaller{\qquad+\sum\nolimits_{j}p_j(t)\sum\nolimits_{i}\big[x_{ij}(t)+\sum\nolimits_ky_{ikj}(t)\big]},
\end{align}
where $c_{ij}(t)$ refers to the cost of transmitting one data sample (i.e., the unit transmission cost) between AM $i$ and MC $j$, $e_{jk}(t)$ refers to the unit transmission cost between MC $j$ and MC $k$, and $p_j(t)$ refers to the unit computing cost of MC $j$. They could be time-varying and inversely proportional to the available resources \cite{pu2018online}. Note that we do not take the global model transmission cost between CN and AMs into account, since it is independent with the centralized decision making (i.e.,  without control variables).

\textbf{Parameter Server}. The parameter server will aggregate the local ML model parameters from all the MCs every $\mathcal{T}$ slots, and then update the global ML model parameters:
\begin{align} \label{C15}
{\bm{\mathcal{W}}(t)=\frac{\sum\nolimits_{j\in\mathcal{M}}|\mathcal{D}_j(t)|\mathcal{W}_j(t)}{|\mathcal{D}(t)|}},
\end{align}
where $t=n\mathcal{T}, n=\{1,2,\dots\}$ and $\mathcal{D}(t)\triangleq\cup_j\mathcal{D}_j(t)$ refers to the overall dataset trained in the framework in slot $t$. Next, the parameter server will synchronize the local ML model parameters of each MC with the updated global ML model parameters (i.e., $\mathcal{W}_j(t)=\bm{\mathcal{W}}(t), \forall j\!\in\!\mathcal{M}$). Note that how to choose the best $\mathcal{T}$ is beyond the scope of this paper, and we set $\mathcal{T}=1$ for simplicity. That is, the parameter server executes the global aggregation every slot. We believe the global aggregation of the in-network training will not lead to great transmission overhead, since the training workers (i.e., MCs) are in the proximity of the parameter server (i.e., CN) and the updated model parameters are of small size. To indicate the convergence degree of the global ML model, we consider the commonly-used mean squared error (MSE) as the loss function. Note that other kinds of loss functions such as linear regression \cite{wang2018edge} are also applicable.

\subsection{Problem Formulation}
Our main purpose is to achieve an online cost-efficient and data skew-aware incremental learning framework in 5G networks. Based on the preceding framework model, we formulate the data scheduling (i.e., data collection and data training) problem as follows:
\begin{align}
& {\mathrm{max}}
& & \mathcal{P}_0\!=\!-\mathsmaller{\mathop{\lim}\limits_{T \to \infty}\frac{1}{T}\sum\nolimits_{t=1}^{T}\mathbb{E}\Big[\mathcal{C}(t)\Big]} \notag\\
& \mathrm{s.\,t.}
& & \text{(2), (3), (5), (6), (7), (8), (10), (11), (13)}\notag\\
&
& & \text{the queue $Q_i(t)$ in (1), $R_{ij}(t)$ in (12) is stable.}\notag\\
& \mathrm{var}
& & \alpha_{ij}(t), z_{jk}(t)\in\{0,1\}, \theta_{ij}(t), x_{ij}(t), y_{ijk}(t)\ge0. \notag
\end{align}
More formally, we introduce the following two time-average constraints to capture the queue stability, since a queue is stable if the average arrival rate is not more than the average departure rate:
\begin{subequations}
\begin{align}
& \mathsmaller{\mathop{\lim}\limits_{T \to \infty}\frac{1}{T}\sum\nolimits_t\mathbb{E}\Big[A_i(t)-\sum\nolimits_{j}\alpha_{ij}(t)\theta_{ij}(t)d_{ij}(t)\Big]\!\le\!0},\label{C16:cons1}\\
& \mathsmaller{\mathop{\lim}\limits_{T \to \infty}\frac{1}{T}\sum\nolimits_t\mathbb{E}\Big[\alpha_{ij}(t)\theta_{ij}(t)d_{ij}(t)\!-\!x_{ij}(t)}\notag\\
& \mathsmaller{\qquad \qquad \quad \qquad \qquad \qquad \qquad \!-\!\sum\nolimits_{k}y_{ijk}(t)\Big]\!\le\!0}.\label{C16:cons2}
\end{align}
\end{subequations}

\textbf{Algorithmic Challenges}: The main challenges of the problem $\mathcal{P}_0$ stem from three aspects. First, it is required to continuously adapt to system dynamics on the fly, since the control variables are temporally coupled in the queues (1), (12) and time-average constraints (10), (11), and besides the future system state $S(t)$ and data arrivals $A_i(t), \forall i$ are hard to predict.
Second, it belongs to the mixed-integer programming, which is difficult to optimally solve with polynomial time. Third, as ML training generally demands for fast convergence, the proposed algorithm should decrease the AM queue backlog in each MC as much as possible (i.e., training more data to accelerate convergence\footnote{Different from many FL based works that can explicitly analyze the trained model convergence with a close-form expression, our framework involves a series of system dynamics (e.g., data arrivals and system state) and requires complicated decision making (i.e., the control variable $x,y,z$ are coupled in many constraints), which is very hard to derive a close-form expression for the convergence analysis. Therefore, we simply consider that the more data samples from diverse AMs are trained indicates the more quickly the global ML model converges. We will justify this argument with dataset-driven evaluations in Section V, and besides we will explore the close-form expression for the convergence analysis in the future work.}.
To sum up, an online algorithm with a provable theoretical performance on both objective and queue backlog as well as polynomial running time is desirable.

\section{Online Algorithm Design}
To address the above challenges, we first derive the Lagrangian dual problem $\mathcal{L}_0$ of the time-correlated $\mathcal{P}_0$, which consists of a series of time-independent per-slot problems and each per-slot problem can be separated into a data collection and a data training subproblem. Then, we exploit the stochastic gradient descent (SGD) technique to devise \emph{DataSche}, an online data scheduling algorithm to solve $\mathcal{L}_0$. As the core building block, we treat those two subproblems in a skew-aware manner and propose optimal algorithms based on novel graph constructions to respectively solve
them. We theoretically analyze the proposed online algorithm and find that there is an undesirable tradeoff between the objective (i.e., framework cost) and the queue backlog (i.e., the number of untrained data samples). To this end, we improve the proposed algorithm with online learning, which can greatly reduce the queue backlog without damaging the objective in theory.

\subsection{DataSche Algorithm}\label{SGD}
Our basic idea for the time-correlated problem $\mathcal{P}_0$ is to lift the time-average constraints to the objective with Lagrange multipliers and iteratively update the multipliers with the stochastic gradients in terms of the time-varying system state and data arrivals in each slot. In this context, we rewrite the constraint (10) and (11) in our framework as follows:
\begin{subequations}
\begin{align}
& \mathsmaller{\mathop{\lim}\limits_{T \to \infty}\frac{1}{T}\sum\nolimits_t\mathbb{E}\Big[\check{\delta}_i\sum\nolimits_l\big[x_{lj}(t)\!+\!\sum\nolimits_k y_{lkj}(t)\big]}\notag\\
& \mathsmaller{\qquad \qquad \qquad \qquad \qquad -x_{ij}(t)\!-\!\sum\nolimits_k y_{ikj}(t)\Big]\!\le\!0},\label{C16:cons3}\\
& \mathsmaller{\mathop{\lim}\limits_{T \to \infty}\frac{1}{T}\sum\nolimits_t\mathbb{E}\Big[x_{ij}(t)\!+\!\sum\nolimits_k y_{ikj}(t)}\notag\\
& \mathsmaller{\qquad \qquad \quad \qquad -\hat{\delta}_i\sum\nolimits_l\big[x_{lj}(t)\!+\!\sum\nolimits_k y_{lkj}(t)\big]\Big]\!\le\!0}\label{C16:cons4}.
\end{align}
\end{subequations}
Then, we denote by\footnote{In this paper, we exploit the bold form of a parameter or variable to represent the finite set of it. For example, $\bm{\mu}(t)\triangleq\{\mu_1(t),\mu_2(t),\dots,\mu_N(t)\}.$} $\bm{\Theta}(t)=\{\bm{\mu}(t),\bm{\eta}(t),\bm{\varphi}(t),\bm{\lambda}(t)\}$ the Lagrange multipliers associated with (\ref{C16:cons1}), (\ref{C16:cons2}), (\ref{C16:cons3}) and (\ref{C16:cons4}), and derive the following Lagrangian dual problem $\mathcal{L}_0$ of problem $\mathcal{P}_0$:
\begin{align}
& \min
& & \mathsmaller{\mathcal{L}_0=\mathop{\lim}\limits_{T \to \infty}\frac{1}{T}\sum\nolimits_t\mathbb{E}\Big[\mathcal{L}(\bm{\Theta}(t))\Big]}\notag\\
&
& & \mathsmaller{\quad \ =\mathop{\lim}\limits_{T \to \infty}\frac{1}{T}\sum\nolimits_t\mathbb{E}\Big[\mathcal{L}_1(\bm{\Theta}(t))\!+\!\mathcal{L}_2(\bm{\Theta}(t))\Big]}\notag\\
& \mathrm{var}
& & \mathsmaller{\mu_i(t), \eta_{ij}(t),\varphi_{ij}(t), \lambda_{ij}(t)\ge0},\notag\\
& \mathrm{where}
& & \mathsmaller{\mathcal{L}_1(\bm{\Theta}(t)) \!=\! \max \ \mathcal{P}_1(\bm{\alpha}(t),\bm{\theta}(t))}\notag\\
&
& & \mathsmaller{\qquad \qquad \quad \  \mathrm{s.\,t.} \ (2), (3)}.\notag\\
&
& & \mathsmaller{\qquad \qquad \quad \  \mathrm{var} \ \, \alpha_{ij}(t)\in\{0,1\}, \theta_{ij}(t)\ge0}.\notag\\
&
& & \mathsmaller{\mathcal{L}_2(\bm{\Theta}(t)) \!=\! \max \ \mathcal{P}_2(\bm{x}(t),\bm{y}(t),\bm{z}(t))}\notag\\
&
& & \mathsmaller{\qquad \qquad \quad \  \mathrm{s.\,t.} \ (5), (6), (7), (8), (13)}.\notag\\
&
& & \mathsmaller{\qquad \qquad \quad \  \mathrm{var} \ \, z_{jk}(t)\in\{0,1\}, x_{ij}(t), y_{ijk}(t)\ge0}.\notag
\end{align}
The specific expressions of $\mathcal{P}_1$ and $\mathcal{P}_2$ are given by
\begin{align}
& \mathsmaller{\mathcal{P}_1\!=\!\sum\nolimits_i\sum\nolimits_j\alpha_{ij}(t)\theta_{ij}(t)d_{ij}(t)\big[\mu_i(t)\!-\!\eta_{ij}(t)\!-\!c_{ij}(t)\big]},\\
& \mathsmaller{\mathcal{P}_2\!=\!\sum\nolimits_{i}\sum\nolimits_{j}\Big(x_{ij}(t)\Big[\!-\!p_j(t)\!+\!\eta_{ij}(t)\!-\!\lambda_{ij}(t)\!+\!\varphi_{ij}(t)}\notag \\
 & \mathsmaller{+\sum\nolimits_{l}[\lambda_{lj}(t)\hat{\delta}_l\!-\!\varphi_{lj}(t)\check{\delta}_l]\Big]\!+\!\sum\nolimits_{k}y_{ikj}(t)\Big[\!-\!p_j(t)\!-\!e_{kj}(t)}\notag\\
& \mathsmaller{+
\eta_{ik}(t)\!-\!\lambda_{ij}(t)\!+\!\varphi_{ij}(t)\!+\!\sum\nolimits_{l}[\lambda_{lj}(t)\hat{\delta}_l-\varphi_{lj}(t)\check{\delta}_l]\Big]\Big)}.
\end{align}
As we can see, the Lagrangian dual problem $\mathcal{L}_0$ consists of a series of time-independent per-slot problems, and each per-slot problem can be further separated into two subproblems: $\mathcal{L}_1$ (data collection) and $\mathcal{L}_2$ (data training).
In this context, instead of directly coping with the primal problem $\mathcal{P}_0$, we design \emph{DataSche}, an efficient online data scheduling algorithm via stochastic gradient descent to solve the dual problem $\mathcal{L}_0$, which can achieve an asymptotical optimum for the problem $\mathcal{P}_0$. This online algorithm mainly consists of two steps:

\begin{itemize}
  \item \textbf{Step 1}. Obtain the system state $S(t)$ and the Lagrange multipliers $\bm{\Theta}(t)$ at the beginning of each slot. Then, derive the optimal data collection decisions  $\bm{\alpha}(t), \bm{\theta}(t)$ by solving the subproblem $\mathcal{L}_1$, and the optimal data training decisions $\bm{x}(t), \bm{y}(t), \bm{z}(t)$ by solving the subproblem $\mathcal{L}_2$.
  \item \textbf{Step 2}. Update the Lagrange multipliers $\bm{\Theta}(t)$ by gradient descent, in terms of the data arrivals $\bm{A}(t)$ and the decisions in Step 1.
\end{itemize}
The specific update rule for $\bm{\Theta}(t)$ is given by
\begin{align}
& \mathsmaller{\mu_i(t+1)\!=\!\Big[\mu_i(t)\!+\!\epsilon\big(A_i(t)\!-\!\sum\nolimits_{j}\alpha_{ij}(t)\theta_{ij}(t)d_{ij}(t)\big)\Big]^+},\notag\\
& \mathsmaller{\eta_{ij}(t+1)\!=\!\Big[\eta_{ij}(t)\!+\!\epsilon\big(\alpha_{ij}(t)\theta_{ij}(t)d_{ij}(t)\!-\!x_{ij}(t)}\notag\\
& \mathsmaller{\qquad \qquad \qquad \qquad \qquad \qquad \qquad \qquad \!-\!\sum\nolimits_{k}y_{ijk}(t)\big)\Big]^+},\notag\\
& \mathsmaller{\varphi_{ij}(t+1)=\Big[\varphi_{ij}(t)\!+\!\epsilon\big(\check{\delta}_i\sum\nolimits_l\big[x_{lj}(t)\!+\!\sum\nolimits_k y_{lkj}(t)\big]}\notag\\
& \mathsmaller{\qquad \qquad \qquad \qquad \qquad \qquad \qquad -x_{ij}(t)\!-\!\sum\nolimits_k y_{ikj}(t)\big)\Big]^+},\notag\\
& \mathsmaller{\lambda_{ij}(t+1)=\Big[\lambda_{ij}(t)\!+\!\epsilon\big(x_{ij}(t)\!+\!\sum\nolimits_k y_{ikj}(t)}\notag\\
& \mathsmaller{\qquad \qquad \quad \qquad \qquad \qquad-\hat{\delta}_i\sum\nolimits_l\big[x_{lj}(t)\!+\!\sum\nolimits_k y_{lkj}(t)\big]\big)\Big]^+},\notag
\end{align}
where $\epsilon$ is the step-size of gradient descent, which is generally given a small value to facilitate the objective optimality.

\textbf{Remark}. We can observe that the backlog of each queue is equivalent to its corresponding Lagrange multiplier over the step-size (e.g., $Q_i(t)=\mu_i(t)/\epsilon$). In addition, although the stochastic gradient descent technique is a general optimization tool \cite{cui2019stochastic, huang2014power, chen2017learn}, the specific forms of two subproblems in Step 1 are unique and accordingly require to design new algorithms.

\subsection{Data Collection}\label{FDC}
The specific expression of the data collection subproblem $\mathcal{L}_1$ is as follows:
\begin{align}
& {\mathrm{max}}
& & \mathsmaller{\sum\nolimits_i\sum\nolimits_j\alpha_{ij}(t)\theta_{ij}(t)d_{ij}(t)\big[\mu_i(t)\!-\!\eta_{ij}(t)\!-\!c_{ij}(t)\big]}\notag \\
& \mathrm{s.\,t.}
& & \mathsmaller{\sum\nolimits_{j}\alpha_{ij}(t)\le1, \quad \sum\nolimits_{i}\theta_{ij}(t)\le1},\notag\\
& \mathrm{var}
& & \alpha_{ij}(t)\in\{0,1\}, \quad \theta_{ij}(t)\ge0.\notag
\end{align}

\textbf{Interpretation}. The expression $\mu_i(t)-\eta_{ij}(t)$ can be interpreted as the backlog gap between the queue $Q_i(t)$ and $R_{ij}(t)$. Intuitively, a larger positive gap indicates that the number of data samples collected from AM $i$ is insufficient in MC $j$, which could lead to the skewed data training over time. The above formulation will implicitly narrow the backlog gap, since MC $j$ is more likely to collect data from AM $i$ if $\mu_i(t)-\eta_{ij}(t)$ is pretty large.

Obviously, $\mathcal{L}_1$ is a mixed integer programming problem, in which $\alpha_{ij}(t), \forall i, \forall j$ can only take binary values. Nevertheless, we can optimally solve it in polynomial time by casting it into a bipartite graph matching problem. Our basic idea is to consider the optimal time allocation when the AM\! --\! MC connections are given, and based on it construct a bipartite graph to derive the optimal AM\! --\! MC connections.

\textbf{Time Allocation of One MC}. When the AM\! --\! MC connections are given (i.e., $\alpha_{ij}(t), \forall i, \forall j$), we can independently consider the time allocation of each MC $j$. To proceed, if we assume there are $N_j(t)$ AMs connected to MC $j$ in slot $t$, then $\mathcal{L}_1$ will degrade to the following problem:
\begin{align} \label{C19}
& {\mathrm{max}}
& & \mathsmaller{\sum\nolimits_{i}\theta_{ij}(t)d_{ij}(t)\big[\mu_i(t)\!-\!\eta_{ij}(t)\!-\!c_{ij}(t)\big]} \\
& \mathrm{s.\,t.}
& & \mathsmaller{\sum\nolimits_{i=1}^{N_j(t)}\theta_{ij}(t)\le1},\notag\\
& \mathrm{var}
& & \theta_{ij}(t)\ge0.\notag
\end{align}
We can easily derive the optimal time allocation $\theta_{i^*j}(t)=1$ for AM $i^*$ satisfying $i^*\!=\!\arg\max d_{ij}(t)$ $\big[\mu_i(t)\!-\!\eta_{ij}(t)\!-\!c_{ij}(t)\big]$, and $\theta_{ij}(t)=0$ for AM $i \neq i^*$. That is, allocate all the time to the AM with the maximum weight.

\textbf{Bipartite Graph Construction}. According to the optimal time allocation of one MC, we construct a bipartite graph $BG\!=\!\{\mathcal{N}, \mathcal{M}, E\}$. $\mathcal{N}$ is the AM set and $\mathcal{M}$ is the MC set. We denote by $e_{ij}$ the edge between AM $i$ and MC $j$ and by $\omega_{ij}(t)$ the edge weight which equals to $d_{ij}(t)\big[\mu_i(t)\!-\!\eta_{ij}(t)\!-\!c_{ij}(t)\big]$ in each slot $t$. Note that the time complexity of this bipartite graph construction is $\mathcal{O}\big(NM\big)$.

In the context of the above bipartite graph construction, we have the following Theorem.

\textbf{Theorem 1}. The optimal solution of $\mathcal{L}_1$ is equivalent to the maximum weight matching on the bipartite graph $BG$.

\begin{IEEEproof}
{We denote by $A$ the optimal objective of $\mathcal{L}_1$ and by $B$ the total weight of the maximum weight matching on the bipartite graph $BG$. If we can prove $A\!\ge\!B$ and meanwhile $B\!\ge\!A$, then Theorem 1 is derived.}

{We first prove $A\!\ge\!B$. Given the maximum weight matching on $BG$, we can derive a series of disjoint subgraph $BG_j, \forall j\in\mathcal{M}$, and each subgraph $BG_j$ includes only one AM denoted by $i^*$ connected to MC $j$. We denote by $B_j$ the total weight of the subgraph $BG_j$ and intuitively we have $B=\sum\nolimits_{j}B_j$ and $B_j=d_{i^*j}(t)\big[\mu_{i^*}(t)\!-\!\eta_{i^*j}(t)\!-\!c_{i^*j}(t)\big]$. In other words, we have $\alpha_{i^*j}(t)=\theta_{i^*j}(t)=1$ and $\alpha_{ij}(t)=\theta_{ij}(t)=0$, $\forall i\neq i^*$. Note that it is a feasible solution of $\mathcal{L}_1$, and accordingly we can have $A\ge B$ since $A$ is the optimal objective of $\mathcal{L}_1$.}

{We next prove $B\!\ge\!A$. Consider the optimal solution of $\mathcal{L}_1$. Given the optimal AM\! --\! MC connections $\alpha_{ij}(t), \forall i, \forall j$, we can partition the bipartite graph $BG$ into a series of disjoint subgraphs $BG'_j, \forall j\in \mathcal{M}$ and each subgraph $BG'_j$ includes some AMs connected to MC $j$ in terms of the optimal time allocation of one MC $\theta_{ij}(t), \forall i, \forall j$. We denote by $B'_j$ the total weight of the subgraph $BG'_j$ and intuitively we have $A=\sum\nolimits_{j}B'_j$. In addition, we denote by $B^*_j$ the maximum weight of the subgraph $BG'_j$ which is attributed to only one edge in terms of the optimal solution of the problem in (20). In this context, we can have $A=\sum\nolimits_{j}B'_j\le\sum\nolimits_{j}B^*_j\le B$, since $B$ is the total weight of the maximum weight matching on the bipartite graph $BG$.} \end{IEEEproof}

According to Theorem 1, we can exploit the classic Hungarian algorithm to derive the maximum weight matching on the bipartite graph $BG$ as the optimal solution of $\mathcal{L}_1$ with the time complexity $\mathcal{O}\big((N\!+\!M)^3\big)$.

\textbf{Algorithmic Limitation}. Nevertheless, the above data collection policy has a clear drawback: \emph{each MC will collect the data samples from only one AM in each time slot}. In other words, only $M$ out of $N$ AMs can upload their data samples per slot. Therefore, this policy will result in a skewed data collection, which could adversely impact the accuracy of local ML model. To this end, we should design a skew-aware data collection policy.

\textbf{Skew-aware Data Collection}. In this case, we instead solve the following skew-aware data collection subproblem $\mathcal{L}'_1$:
\begin{align}
& {\mathrm{max}}
& & \mathsmaller{\sum\nolimits_i\log\Big(\sum\nolimits_j\alpha_{ij}(t)\theta_{ij}(t)d_{ij}(t)\big[\mu_i(t)\!-\!\eta_{ij}(t)\!-\!c_{ij}(t)\big]\Big)}\notag \\
& \mathrm{s.\,t.}
& & \mathsmaller{\sum\nolimits_{j}\alpha_{ij}(t)\le1},\notag\\
&
& & \mathsmaller{\sum\nolimits_{i}\theta_{ij}(t)\le1},\notag\\
& \mathrm{var}
& & \alpha_{ij}(t)\in\{0,1\}, \theta_{ij}(t)\ge0.\notag
\end{align}
{We exploit the logarithmic operator in the objective to ensure each MC will collect the data samples from multiple AMs in a proportional fairness manner. Note that such a logarithmic utility function is commonly used in the proportional-fair rate control for communication networks \cite{neely2010stochastic}.}

We can exploit the same idea to optimally solve $\mathcal{L}'_1$. Specifically, we first consider the time allocation of one MC. When the AM\! --\! MC connections are given (i.e., $\alpha_{ij}(t), \forall i, \forall j$), $\mathcal{L}'_1$ will degrade to the following problem:
\begin{align}
& {\mathrm{max}}
& & \mathsmaller{\sum\nolimits_{i}\log\Big(\theta_{ij}(t)d_{ij}(t)\big[\mu_i(t)\!-\!\eta_{ij}(t)\!-\!c_{ij}(t)\big]\Big)} \\
& \mathrm{s.\,t.}
& & \mathsmaller{\sum\nolimits_{i=1}^{N_j(t)}\theta_{ij}(t)\le1},\notag\\
& \mathrm{var}
& & \theta_{ij}(t)\ge0.\notag
\end{align}
Different from the optimal solution of the problem in (20), the optimal time allocation $\theta_{ij}(t)\!=\!1/N_j(t), {\forall i=1,\dots,N_j(t)}$, since $\log(a_1\theta_1)\!+\!\log(a_2\theta_2)=\log(a_1a_2)\!+\!\log(\theta_1\theta_2)$ and the inequality $\theta_1\theta_2\cdots\theta_n\le(\frac{\theta_1+\theta_2+\dots+\theta_n}{n})^2\le\frac{1}{n^2}$ takes equality only if $\theta_1=\theta_2=\dots=\theta_n=\frac{1}{n}$. That is, evenly allocate the time to the connected AMs.

\begin{figure}[thb]
\centering
\includegraphics[height=1.4in]{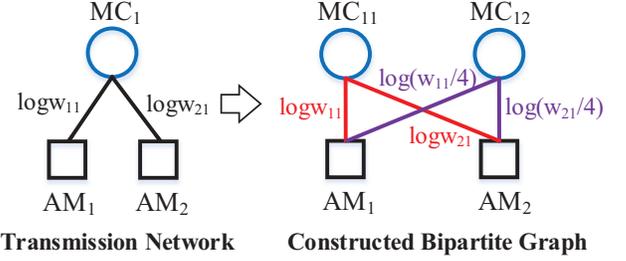}
\caption{Example of bipartite graph construction.}
\label{fig:bipartite2}
\end{figure}

To proceed, we construct a bipartite graph $BG'\!=\!\{U, V, E\}$. $U=\mathcal{N}$ refers to the AM set and we introduce $N$ virtual MCs denoted by $\{v_{j1}, v_{j2},\dots, v_{jN}\}$ for each MC $j\in\mathcal{M}$ in the set $V$ (i.e., $|V|=NM$). We denote by $e_{ij}^n$ the edge between AM $i$ and virtual MC $v_{jn}$ and by $\omega_{ij}^n(t)$ the edge weight which equals to $\log\big((n-1)^{n-1}{\text{w}}_{ij}(t)/n^n\big)$, where ${\text{w}}_{ij}(t)\triangleq d_{ij}(t)\big[\mu_i(t)\!-\!\eta_{ij}(t)\!-\!c_{ij}(t)\big]$ in each slot $t$. The time complexity of this graph construction is $\mathcal{O}\big(N^2M\big)$. Consider a simple AM\! --\! MC transmission network as shown in Fig. \ref{fig:bipartite2}. As it contains two AMs (i.e., AM$_{1}$ and AM$_{2}$) and one MC, we should introduce two virtual MCs (i.e., MC$_{11}$ and MC$_{12}$) in the constructed bipartite graph, and add an edge between each pair of AM and virtual MC with the specific weight discussed above. For example, $\omega_{11}^1=\log{\text{w}}_{11}$ and $\omega_{21}^2=\log({\text{w}}_{21}/4)$.

The motivation of such a bipartite graph construction illustrated with Fig. \ref{fig:bipartite2} is that if only AM$_{1}$ is connected to MC$_{1}$, $\theta_{11}=1$ in terms of the optimal time allocation of one MC, and therefore the optimal objective in (21) is $\log{\text{w}}_{11}$. Similarly, it is $\log({\text{w}}_{11}/2)+\log({\text{w}}_{21}/2)$ if both AM$_{1}$ and AM$_{2}$ are connected to MC$_{1}$. We can find that the marginal objective gain for adding AM$_{2}$ is $\log({\text{w}}_{11}/2)+\log({\text{w}}_{21}/2)-\log{\text{w}}_{11}=\log({\text{w}}_{21}/4)$, exactly the edge weight $\omega_{21}^2$ between AM$_2$ and MC$_{12}$. In this context, we can prove that the optimal objective in (21) when there are $n$ AMs connected to MC $j$ equals to the total weight of matching AM $i$ to the virtual MC $v_{ji}, \forall i=1,2,\dots,n$.
\begin{IEEEproof}
{The optimal objective in (21) when there are $n$ AMs connected to MC $j$ is $\sum\nolimits_{i=1}^{n}\log({\text{w}}_{ij}/n)$ in terms of the optimal time allocation of one MC. The total weight of matching AM $i$ to the virtual MC $v_{ji}, \forall i=1,2,\dots,n$ is $\sum\nolimits_{i=1}^{n}\omega_{ij}^i$ which by definition equals to $\sum\nolimits_{i=1}^{n}\log\big((i-1)^{i-1}{\text{w}}_{ij}/i^i\big)$. According to the property of the logarithmic operator, we have
\begin{align}
&
& & \mathsmaller{\sum\nolimits_{i=1}^{n}\log\big((i-1)^{i-1}{\text{w}}_{ij}/i^i\big)}\notag\\
&
& & \mathsmaller{=\sum\nolimits_{i=1}^{n}\log\big(\text{w}_{ij}\big)+\sum\nolimits_{i=1}^{n}\log\big((i-1)^{i-1}/i^i\big)}\notag\\
&
& & \mathsmaller{=\sum\nolimits_{i=1}^{n}\log\big(\text{w}_{ij}\big)+\log\big(\frac{1}{1}\times\frac{1}{4}\times\frac{4}{9}\times\dots\times\frac{(n-1)^{n-1}}{n^n}\big)}\notag\\
&
& & \mathsmaller{=\sum\nolimits_{i=1}^{n}\log\big(\text{w}_{ij}\big)+\log\big(1/n^n\big)=\sum\nolimits_{i=1}^{n}\log({\text{w}}_{ij}/n)},\notag
\end{align}
which confirms our argument.}
\end{IEEEproof}

In terms of the above bipartite graph construction, we can have the following Theorem.

\textbf{Theorem 2}. The optimal solution of $\mathcal{L}'_1$ is equivalent to the maximum weight matching on the bipartite graph $BG'$.

{The proof is very similar to that of Theorem 1, and therefore we leave it in our online technical report \cite{report}.} According to Theorem 2, we can exploit the classic Hungarian algorithm to derive the maximum weight matching on the bipartite graph $BG'$ as the optimal solution of $\mathcal{L}'_1$ with the time complexity $\mathcal{O}\big((N+NM)^3\big)$. As a conclusion, although its time complexity is dominated by $\mathcal{O}\big(N^3M^3\big)$, the skew-aware data collection enables each MC to collect the data samples from multiple AMs in a proportional fairness manner per slot, which facilitates the sample evenness.

\subsection{Data Training}\label{EDT}
The specific expression of the data training subproblem $\mathcal{L}_2$ is as follows:
\begin{align}
& {\mathrm{max}}
& & \mathsmaller{\sum\nolimits_{i}\sum\nolimits_{j}\Big(x_{ij}(t)\Big[\!-\!p_j(t)\!+\!\eta_{ij}(t)\!-\!\lambda_{ij}(t)\!+\!\varphi_{ij}(t)}\notag \\
&
& & \mathsmaller{\!+\!\sum\nolimits_{l}[\lambda_{lj}(t)\hat{\delta}_l\!-\!\varphi_{lj}(t)\check{\delta}_l]\Big]\!+\!\sum\nolimits_{k}y_{ikj}(t)\Big[\!-\!p_j(t)\!-\!e_{kj}(t)}\notag\\
&
& & \mathsmaller{\!+
\eta_{ik}(t)\!-\!\lambda_{ij}(t)\!+\!\varphi_{ij}(t)\!+\!\sum\nolimits_{l}[\lambda_{lj}(t)\hat{\delta}_l-\varphi_{lj}(t)\check{\delta}_l]\Big]\Big)}\notag\\
& \mathrm{s.\,t.}
& &
(5), (6), (7), (8), (13).\notag\\
& \mathrm{var}
& & z_{jk}(t)\in\{0,1\}, x_{ij}(t), y_{ikj}(t)\ge0.\notag
\end{align}

\textbf{Interpretation}. The expression of $\eta_{ik}(t)-\eta_{ij}(t)$ refers to the main difference between the weights of $y_{ikj}(t)$ and $x_{ij}(t)$, which also indicates the backlog gap between the queue $R_{ik}(t)$ and $R_{ij}(t)$. The above formulation implicitly shows that
if MC $j$ lacks some data samples from AM $i$, then it is more likely to request some data from another MC $k$ with sufficient data samples (i.e., a larger $\eta_{ik}(t)-\eta_{ij}(t)$). Such a data offloading contributes to achieving a good sample evenness from the per-slot perspective. In addition, the expression $\varphi_{ij}(t)-\lambda_{ij}(t)$ indicates the skew degree of the trained data from AM $i$ in MC $j$ from the long-term perspective.
That is, a positive value implies the number of trained data from AM $i$ to date is less than the expected one. The above formulation implicitly decreases this value, since MC $j$ will train more data from AM $i$ if $\varphi_{ij}(t)\!-\!\lambda_{ij}(t)$ is pretty large, which reveals the long-term skew amendment.

$\mathcal{L}_2$ is also a mixed integer program, in which $z_{jk}(t), \forall j, \forall k$ can only take binary values. Nevertheless, we can optimally solve it in polynomial time by casting it into a graph matching problem. In terms of constraint (5), we can know that each MC has either no connection (i.e., local training without MC cooperation) or exactly one connection to another MC (i.e., local training with MC cooperation) under any a data training policy. Therefore, our basic idea is to study the optimal data training decisions in these two cases, and then construct a graph to derive the optimal MC\! --\! MC connections.

\textbf{Local Training Without MC Cooperation}. In this case, we can independently consider the local training for each MC $j$, and hence we require to solve the following problem:
\begin{align}
& {\mathrm{max}} \quad \ \mathsmaller{\sum\nolimits_{i}\beta_{ij}(t)x_{ij}(t)}\\
& \mathrm{s.\,t.}\quad \ \ \mathsmaller{\sum\nolimits_{i}x_{ij}(t)\rho\le f_j(t)},\notag\\
& \qquad \quad \ \mathsmaller{x_{ij}(t)\le R_{ij}(t),}\notag\\
& \mathrm{var} \quad \ \ \ x_{ij}(t)\ge0,\notag
\end{align}
where $\beta_{ij}(t)$ is the weight of $x_{ij}(t)$ in $\mathcal{L}_2$.
As it belongs to the classic convex optimization, it can be efficiently solved by many mature algorithms such as interior point method.

\textbf{Local Training With MC Cooperation}. In this case, we can independently consider the local training for each pair of MC $j$ and MC $k$, and hence we require to solve the problem:
\begin{align}
& {\mathrm{max}}
& & \mathsmaller{\sum\nolimits_{i}\big(\beta_{ij}(t)x_{ij}(t)\!+\!\gamma_{ikj}(t)y_{ikj}(t)\big)}\notag\\
&
& & \mathsmaller{\qquad+\sum\nolimits_{i}\big(\beta_{ik}(t)x_{ik}(t)\!+\!\gamma_{ijk}(t)y_{ijk}(t)\big)}\\
& \mathrm{s.\,t.}
& &
\mathsmaller{\sum\nolimits_{i}[y_{ijk}(t)+y_{ikj}(t)]\le D_{jk}(t)}, \notag\\
&
& & \mathsmaller{\sum\nolimits_{i}[x_{ij}(t)+y_{ikj}(t)]\rho\le f_j(t)},\notag\\
&
& & \mathsmaller{\sum\nolimits_{i}[x_{ik}(t)+y_{ijk}(t)]\rho\le f_k(t)},\notag\\
&
& & \mathsmaller{x_{ij}(t)+y_{ijk}(t)\le R_{ij}(t)},\notag\\
&
& & \mathsmaller{x_{ik}(t)+y_{ikj}(t)\le R_{ik}(t)},\notag\\
& \mathrm{var}
& & x_{ij}(t), y_{ikj}(t)\ge0,\notag
\end{align}
where $\gamma_{ikj}(t)$ is the weight of $y_{ikj}(t)$ in $\mathcal{L}_2$. It is also a convex optimization problem, and hence can be efficiently solved.

\textbf{Graph Construction}. In terms of the optimal decisions in the above two cases, we construct a graph $G\!=\!\{U', E'\}$. Specifically, we introduce a virtual MC denoted by $j'$ for each MC $j\in\mathcal{M}$ in the set $U'$ (i.e., $|U'|=2M$). We denote by $e_{jj'}$ the edge between MC $j$ and the virtual MC $j'$ with the weight equaling to the optimal objective of the problem in (22). Similarly, we denote by $e_{jk}$ the edge between MC $j$ and MC $k$ with the weight equaling to the optimal objective of the problem in (23). Note that the time complexity of this graph construction is $\mathcal{O}\big(M^2\big)$.

\begin{figure}[thb]
\centering
\includegraphics[height=1.5in]{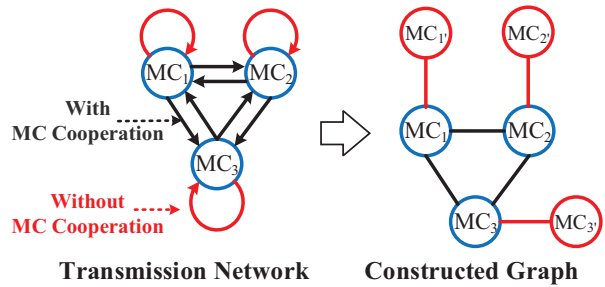}
\caption{Example of graph construction.}
\label{fig:graph}
\end{figure}

Consider a simple MC\! --\! MC transmission network as shown in Fig. \ref{fig:graph}, where a self-loop indicates the local training without MC cooperation and an arrow between two MCs refers to the local training of the ending point assisted by the starting point. As this case contains three MCs, we should introduce three virtual MCs (i.e., MC$_{1'}$, MC$_{2'}$ and MC$_{3'}$) in the constructed graph, and add an edge between each pair of MCs with the specific weight discussed above.

In the context of the above graph construction, we have the following Theorem.

\textbf{Theorem 3}. The optimal solution of $\mathcal{L}_2$ is equivalent to the maximum weight matching on the graph $G$.

The proof is very similar to that of Theorem 1, and therefore we leave it in our online technical report \cite{report}. According to Theorem 3, we can exploit the Edmonds' Blossom algorithm to derive the maximum weight matching on the graph $G$ as the optimal solution of $\mathcal{L}_2$ with the time complexity $\mathcal{O}\big(M^3\big)$.

\textbf{Algorithmic Limitation}. Nevertheless, the above data training policy has an implicit drawback: \emph{each MC will train more data samples from the AM queues with the larger weights in $\mathcal{L}_2$ (i.e., $\beta_{ij}(t)$ and $\gamma_{ikj}(t)$) in each time slot}.  Therefore, this policy will lead to the skewed data training per slot, which could adversely impact the accuracy of local
ML model. To this end, we should design a skew-aware data training policy.

\textbf{Skew-aware Data Training}. In this case, we instead solve the following skew-aware data training subproblem $\mathcal{L}'_2$:
\begin{align}
& {\mathrm{max}}
& & \mathsmaller{\sum\nolimits_{i}\sum\nolimits_{j}\log\Big(x_{ij}(t)\Big[\!-\!p_j(t)\!+\!\eta_{ij}(t)\!-\!\lambda_{ij}(t)\!+\!\varphi_{ij}(t)}\notag \\
&
& & \mathsmaller{\!+\!\sum\nolimits_{l}[\lambda_{lj}(t)\hat{\delta}_l\!-\!\varphi_{lj}(t)\check{\delta}_l]\Big]\!+\!\sum\nolimits_{k}y_{ikj}(t)\Big[\!-\!p_j(t)\!-\!e_{kj}(t)}\notag\\
&
& & \mathsmaller{\!+
\eta_{ik}(t)\!-\!\lambda_{ij}(t)\!+\!\varphi_{ij}(t)\!+\!\sum\nolimits_{l}[\lambda_{lj}(t)\hat{\delta}_l-\varphi_{lj}(t)\check{\delta}_l]\Big]\Big)}\notag\\
& \mathrm{s.\,t.}
& &
(5), (6), (7), (8), (13).\notag\\
& \mathrm{var}
& & z_{jk}(t)\in\{0,1\}, x_{ij}(t), y_{ikj}(t)\ge0.\notag
\end{align}
Similar to the skew-aware data collection, we also exploit the logarithmic operator in the objective to ensure each MC will train the data samples from multiple AM queues in a proportional fairness manner. In addition, the logarithmic operator does not impact the concavity of the objective, and therefore we can exploit the same idea to optimally solve $\mathcal{L}'_2$ (i.e., the only difference is the objective in (22) and (23) involving the logarithmic operator).

As a conclusion, the skew-aware data training enables each MC to train the data samples from multiple AM queues in a proportional fairness manner
per slot, which facilitates the sample evenness. From now on, we consider the \emph{DataSche} algorithm will exploit the skew-aware data collection and skew-aware data training. That is, it will derive the optimal skew-aware data collection and training decisions by solving the subproblem $\mathcal{L}'_1$ and $\mathcal{L}'_2$ in the algorithm step 1.

\subsection{Theoretical Analysis}\label{Performance}
\textbf{Algorithm Complexity}. As the proposed \emph{DataSche} algorithm consists of the skew-aware data collection, the skew-aware data training and the update of the Lagrange multipliers, its time complexity is $\mathcal{O}\big(N^3M^3\big)+\mathcal{O}\big(M^3\big)+\mathcal{O}\big(NM\big)$, dominated by the complexity of skew-aware data collection (i.e, $\mathcal{O}\big(N^3M^3\big))$. We consider it could be acceptable in practice, since our in-network training framework generally has a reasonably long time slot (e.g., at the timescale of tens of minutes) to facilitate model training, and accordingly is less sensitive to the running time of \emph{DataSche} algorithm. In addition, the number of MCs and AMs are limited in 5G networks (e.g., MC sites: 10$\sim$50) \cite{MetroHaul, Cloud5G}. Besides, as the subproblems $L'_1$
and $L'_2$ are independent with each other, we could make decisions on data collection and data training in parallel to accelerate the algorithm execution. Note that we will evaluate the running time of \emph{DataSche} algorithm with extensive experiments in Section V.

\textbf{Performance}. We introduce the following Theorem to indicate the \emph{DataSche} performance.

\textbf{Theorem 4}. Suppose the expected data generation rate of each AM (i.e., $\zeta$) is strictly within the training capacity region $\Lambda$ of our framework, and hence there definitely exists an offline optimal data scheduling policy \{$\Phi^*(t), \forall t$\} for the problem $\mathcal{P}_0$. Denote by $\mathcal{C}(\Phi^*(t))$ the per-slot objective of $\mathcal{P}_0$ produced by the offline policy and by $\mathcal{C}(\Phi(t))$ the per-slot objective of $\mathcal{P}_0$ produced by the \emph{DataSche} algorithm \{$\Phi(t), \forall t$\}. Then, we have the following results:
\begin{align}
& {\mathop{\lim}\limits_{T \to \infty}\frac{\sum\nolimits_{t}\mathbb{E}\big[\mathcal{C}(\Phi(t))\big]}{T}}\!\le\!\frac{\sum\nolimits_{t}\mathcal{C}(\Phi^*(t))}{T}\!+\!\mathcal{O}\big(\varepsilon\big),\\
& {\mathop{\lim}\limits_{T \to \infty}\frac{\sum\nolimits_{t}\mathbb{E}\big[\sum\nolimits_{i}Q_i(t)+\sum\nolimits_{i}\sum\nolimits_{j}R_{ij}(t)\big]}{T}}=\mathcal{O}\big(1/\varepsilon\big),
\end{align}
where $\varepsilon$ is the step-size for updating the Lagrange multipliers in the \emph{DataSche} algorithm as mentioned in Section \ref{SGD}. We can resort to Lyapunov drift-plus-penalty \cite{neely2010stochastic} and weak duality to prove this Theorem. Since the detailed proof is tedious, we leave it in our online technical report \cite{report}.

\textbf{Interpretation and Limitation}. The above results are in accordance with the intuition. Specifically, the inequality (24) indicates that the \emph{DataSche} algorithm achieves an asymptotical optimum (i.e., $\varepsilon\!\to\!0$, a small step-size of gradient descent) for the problem $\mathcal{P}_0$. In other words, the Lagrange multipliers $\bm{\Theta}(t)$ with a small $\varepsilon$ will converge to the neighborhood of the optimum ones $\bm{\Theta}^*$ in the steady state (i.e., $\bm{\Theta}(t) \!\to\! \bm{\Theta}^*+\mathcal{O}\big(1\big)$). Since the backlog of each queue is equivalent to its corresponding Lagrange multiplier over the step-size, we can have that $\bm{\Theta}(t)/\varepsilon\!\to\! \bm{\Theta}^*/\varepsilon+\mathcal{O}\big(1/\varepsilon\big)$, which is in accordance with the inequality (25). That is, the sum of queue backlog is inversely proportional to the step-size $\varepsilon$.

Despite the above results do not give expressions to tight theoretical bounds, they reveal a \big[$\mathcal{O}\big(\varepsilon\big), \mathcal{O}\big(1/\varepsilon\big)$\big] tradeoff between the objective and queue backlog. In practice, we are more likely to give the step-size $\varepsilon$ a small value to facilitate the objective optimality, which inevitably results in a large queue backlog and consequently prolongs the convergence time of global ML model.

\subsection{Learning-aid DataSche Algorithm}
To overcome the limitation of the \emph{DataSche} algorithm, we resort to the dual learning framework \cite{huang2014power, chen2017learn} to improve our algorithm.
The basic idea is to introduce the empirical Lagrange multipliers $\bm{\Theta}'(t)$ to incrementally learn the network state distribution while adapting data collection and training decisions with the learning-aid Lagrange multipliers $\bm{\tilde{\Theta}}(t)$. Here, $\bm{\tilde{\Theta}}(t)\!=\!\bm{\Theta}(t)\!+\!\bm{\Theta}'(t)\!-\!\bm{\pi}, \forall t$ and $\bm{\pi}$ is a dedicated parameter (i.e., $\sqrt{\varepsilon}\log^2(\varepsilon)$ in \cite{huang2014power, chen2017learn}) to control the ``distance" between $\bm{\tilde{\Theta}}(t)$ and the optimal one (i.e., $\bm{\tilde{\Theta}}(t) \!\to\! \bm{\Theta}^*\!+\!\mathcal{O}\big(\bm{\pi}\big)$) in the steady state. The specific steps of the \emph{Learning-aid DataSche} algorithm are given as follows:
\begin{itemize}
  \item \textbf{Step 1}. Obtain the system state $S(t)$ and the learning-aid Lagrange multipliers $\bm{\tilde{\Theta}}(t)$ in each slot. Derive the optimal skew-aware data scheduling decision $\Phi^+(t)$ by solving the subproblem $\mathcal{L}'_1$ and $\mathcal{L}'_2$ with $\bm{\tilde{\Theta}}(t)$.
  \item \textbf{Step 2}. Update the Lagrange multipliers $\bm{\Theta}(t)$ by gradient descent as given in Section \ref{SGD}, in terms of the data arrivals $\bm{A}(t)$ and the decision $\Phi^+(t)$ in Step 1.
  \item \textbf{Step 3}. Obtain the system state $S(t)$ and the empirical Lagrange multipliers $\bm{\Theta}'(t)$ in each slot. Derive the optimal skew-aware data scheduling decision $\Phi'(t)$ by solving the subproblem $\mathcal{L}'_1$ and $\mathcal{L}'_2$ with $\bm{\Theta}'(t)$.
  \item \textbf{Step 4}. Update the empirical Lagrange multipliers $\bm{\Theta}'(t)$ via gradient descent, in terms of the data arrivals $\bm{A}(t)$ and the decision $\Phi'(t)$ in Step 3. Briefly, take $\mu'_i(t)\in\bm{\Theta}'(t), \forall i$ as an example. The update rule is
      \begin{align}
      \mathsmaller{\mu'_i(t\!+\!1)\!=\!\big[\mu'_i(t)\!+\!\sigma(t)\big(A_i(t)\!-\!\sum\nolimits_{j}\alpha'_{ij}(t)\theta'_{ij}(t)d_{ij}(t)\big)\big]^+,}\notag
      \end{align}
      where $\sigma(t)$ is a proper diminishing step-size \cite{chen2017learn}.
  \item \textbf{Step 5}. Update the learning-aid Lagrange multipliers $\bm{\tilde{\Theta}}(t)$, in terms of the empirical Lagrange multipliers $\bm{\Theta}'(t)$ and the Lagrange multipliers $\bm{\Theta}(t)$ as follows:
      \begin{align}
      \mathsmaller{\bm{\tilde{\Theta}}(t)\!=\!\bm{\Theta}(t)\!+\!\bm{\Theta}'(t)\!-\!\bm{\pi}.}\notag
      \end{align}
\end{itemize}

\textbf{Performance}: We can leverage the similar proof in \cite{chen2017learn} to derive the performance of the learning-aid algorithm:
\begin{align}
& {\mathop{\lim}\limits_{T \to \infty}\frac{\sum\nolimits_{t}\mathbb{E}\big[\mathcal{C}(\Phi^+(t))\big]}{T}}\!\le\!\frac{\sum\nolimits_{t}\mathcal{C}(\Phi^*(t))}{T}\!+\!\mathcal{O}\big(\varepsilon\big),\notag\\
& {\mathop{\lim}\limits_{T \to \infty}\frac{\sum\nolimits_{t}\mathbb{E}\big[\sum\nolimits_{i}Q^+_i(t)\!+\!\sum\nolimits_{i}\sum\nolimits_{j}R^+_{ij}(t)\big]}{T}}\!=\! \mathcal{O}\big(\frac{\log^2(\varepsilon)}{\sqrt{\varepsilon}}\big).\notag
\end{align}

We observe that the \emph{Learning-aid DataSche} algorithm also keeps an $\mathcal{O}\big(\varepsilon\big)$ gap compared with the offline optimum, while markedly reducing the queue backlog compared with the \emph{DataSche} algorithm especially when the value of step-size $\varepsilon$ is small. For example, the ratio of the bound of queue backlog between the learning-aid and the original algorithm is $40/100=0.4$ when $\varepsilon=0.01$. As a conclusion, with a moderately small step-size $\varepsilon$, the learning-aid online algorithm will achieve a near-optimal framework cost and a relatively fast training convergence (i.e., training more data samples).
\begin{figure*}[tt]
\centering
\subfigure[Community detection]{
\includegraphics[height=1.62in]{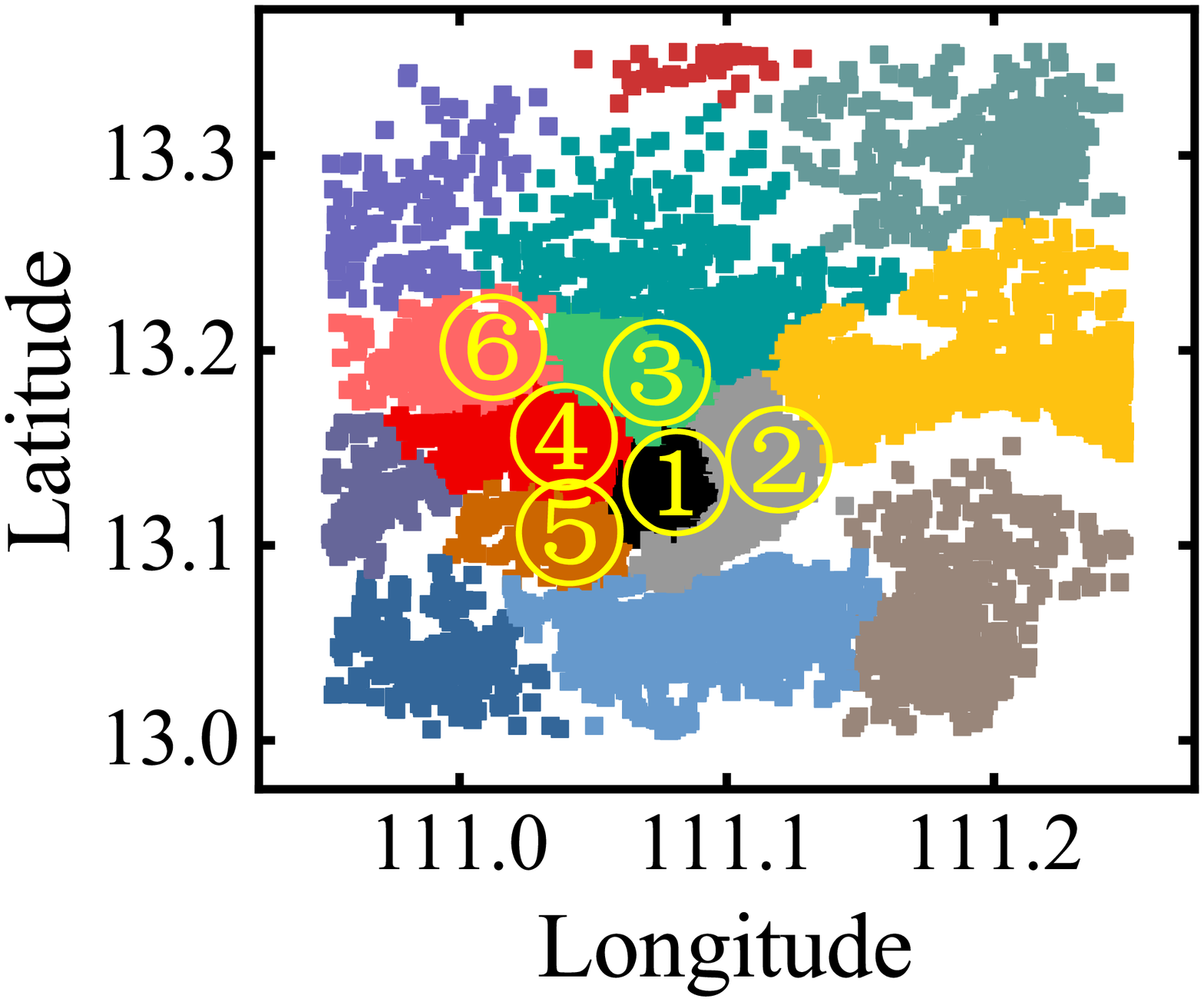}}
\subfigure[Traffic distribution]{
\includegraphics[height=1.5in]{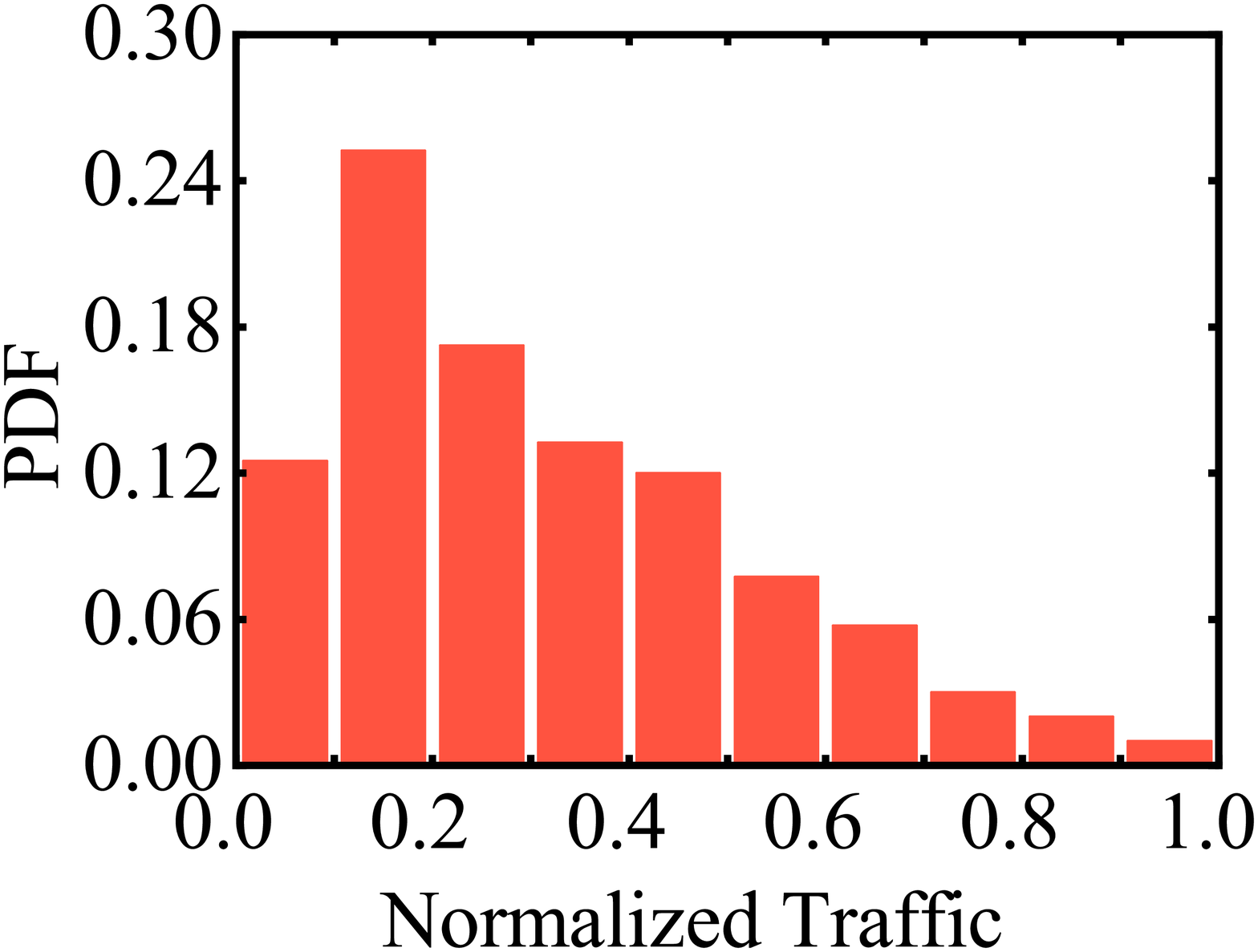}}
\subfigure[Workload distribution]{
\includegraphics[height=1.5in]{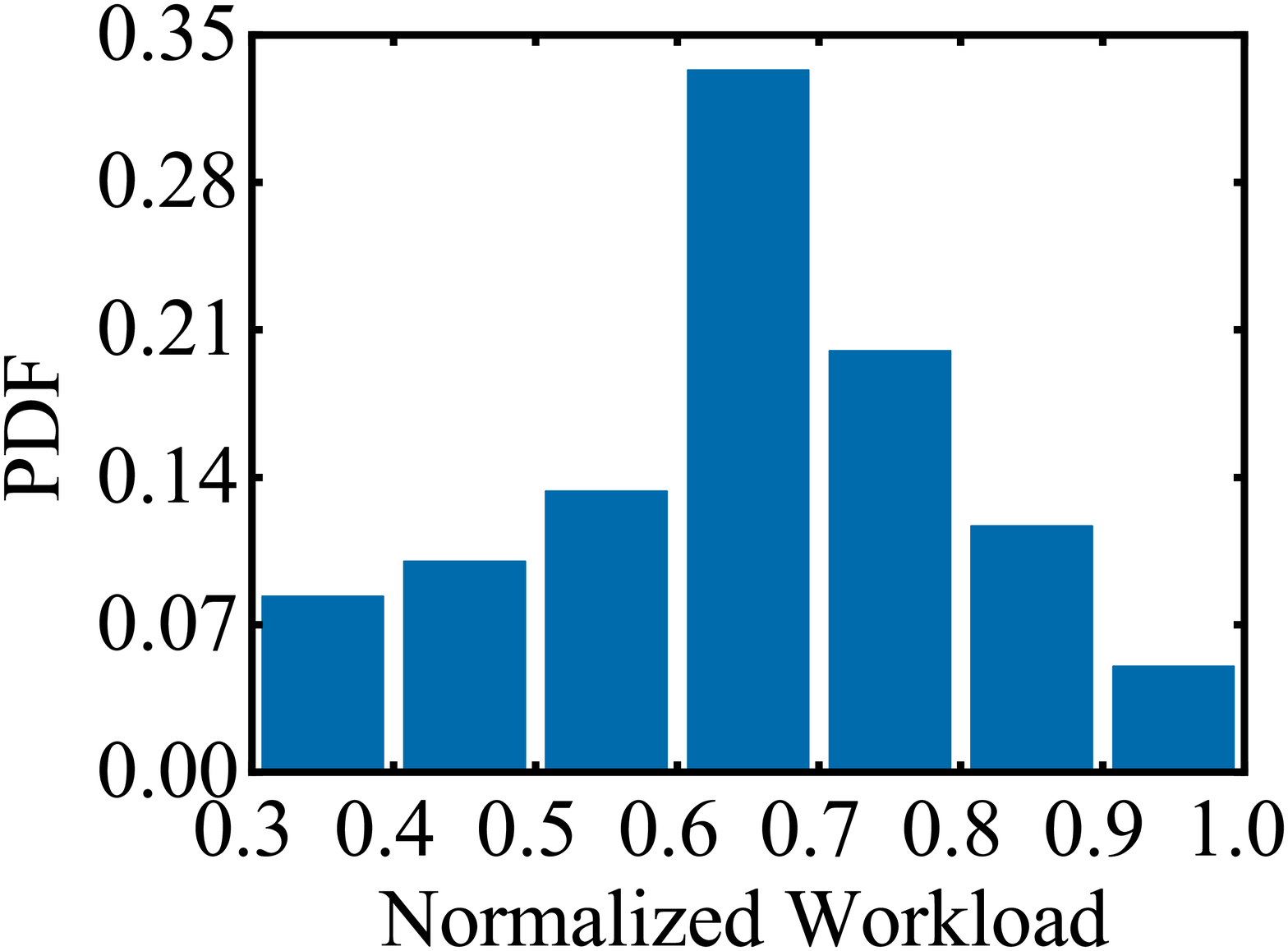}}
\caption{Dataset analysis for evaluation.}
\label{fig:motivation}
\end{figure*}

\section{Performance Evaluation}
In this section, we will evaluate the proposed framework with a small-scale testbed and large-scale simulations.

\subsection{Testbed Implementation}\label{Implementation}
\quad Our testbed is implemented on a software-defined Ethernet LAN consisting of 10 PCs: 1 parameter server (Open vSwitch controller), 3 MCs and 6 AMs. The server and MCs run on Ubuntu 16.04 installed with TensorFlow 2.0.0. One MC has 8 CPU cores@3.0GHz, and the other two have 2 CPU cores@3.0GHz. The MCs will cooperatively train an ML model (i.e., a LSTM network: LSTM(256)+LSTM(256)+Dropout(0.2)+Dense(1)) for cellular traffic prediction with the realistic cellular traffic dataset \cite{Dataset_traffic} (i.e., network prediction function). As for the data generation in AMs, we first build a base station graph where a base station refers to a vertex, and the edge weight refers to the distance between the corresponding base stations in the dataset. Then, we exploit the multilevel community detection algorithm implemented in python igraph library to derive 14 communities as shown in Fig. \ref{fig:motivation}(a). Then, we assign six communities at the center to the AMs. At last, each AM will randomly select 90\% base stations from its assigned community and exploit their traffic records to generate data samples (i.e., the input vector is four time-consecutive traffic records of one base station and the desired scalar output is the next traffic record of that base station). The traffic records of the rest 10\% base stations are used for sample testing. The expected data generation rate $\zeta=500$ with the dynamics following 0\! --\! 1 uniform distribution. The time slot is set to 2 minutes and the number of slots is 60 (i.e., 2 hours).

As for the AM\! --\! MC transmission capacity, we assign a baseline value from \{50, 100\}kbps and provide a real measured traffic distribution to simulate its per-slot dynamics. Briefly, for each AM\! --\! MC connection, we randomly select 100 base stations in the cellular traffic dataset \cite{Dataset_traffic}, and take the total traffic every 2 hours of each base station as a sample. Then, we normalize the derived 9600 samples with the maximum sample value and depict the traffic distribution as shown in Fig. \ref{fig:motivation}(b). In this case,
we can derive the transmission capacity as the baseline$\times$(1 $-$ a randomized normalized traffic per slot)$\times$slot length. The MC\! --\! MC transmission capacity can be set in a similar way (baseline is 300kbps). In the testbed, we will configure the per-slot transmission capacity between PCs with Open vSwitch.

As for the MC computing capacity, the baseline value of each MC is its maximum CPU capacity and the per-slot dynamics is captured by a real measured workload distribution in terms of the Google cluster data trace \cite{google}. Briefly, the dataset records the number of CPU cores utilized by each program every 5 minutes over 29 days. We simply sum up the total number of cores used by all program every 5 minutes as a sample, and randomly select consecutive 5 \!--\! day records (i.e., 1440 samples) for each MC to generate the normalized workload distribution as shown in Fig. \ref{fig:motivation}(c). In this case, we can derive the computing capacity as the baseline$\times$(1 $-$ a randomized normalized workload per slot)$\times$slot length. In the testbed, we will create a program \cite{CPUCOM} to control the CPU usage to achieve the desired computing capacity.

As for the unit computing resource consumption $\rho$, we respectively train 500, 5000, 10000, 20000, 50000 data samples in a MC with 8 CPU cores@3.0GHz, record the CPU running time and CPU usage for the sample training with the top command in Linux, and exploit linear regression to derive that $\rho=1.3\times10^{9}$ cycles. In addition, we simply set the baseline value of the unit cost $c_{ij}(t)=300$, $e_{jk}(t)=50$, $p_j(t)=150, \forall i, \forall j, \forall k$ with the dynamics following 0 \!-- \!1 uniform distribution. Since $1/N=1/6$ (i.e., 6 AMs), we set the tolerance of data skew $\delta$ to 0.02. Unless otherwise noted, the step-size $\varepsilon$ is set to 0.1. All these values are stored in the parameter server. The basic flow of our testbed will be provided in our online technical report \cite{report} for reference.

\subsection{Testbed Evaluation}
\quad Next, we conduct the testbed evaluation by answering the following three questions.

\textbf{(1) Will the skew-aware data collection, skew-aware data training and long-term skew amendment take effect on averting the skewed data training?}
\begin{figure*}[tt]
\begin{minipage}[t]{0.33\linewidth}
\centering
\includegraphics[height=4.2cm]{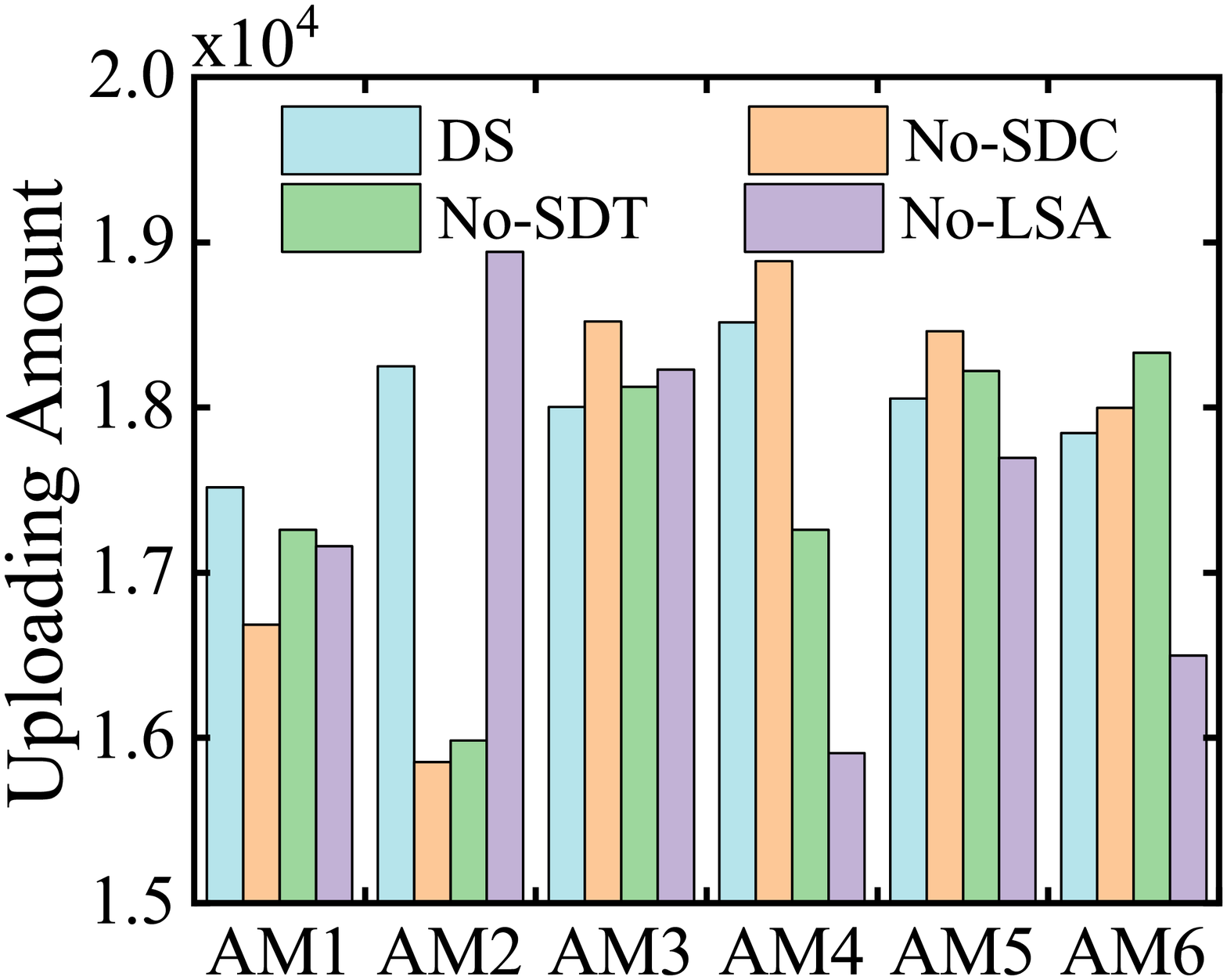}\caption{Data uploading amount.}
\end{minipage}%
\begin{minipage}[t]{0.33\linewidth}
\centering
\includegraphics[height=4.2cm]{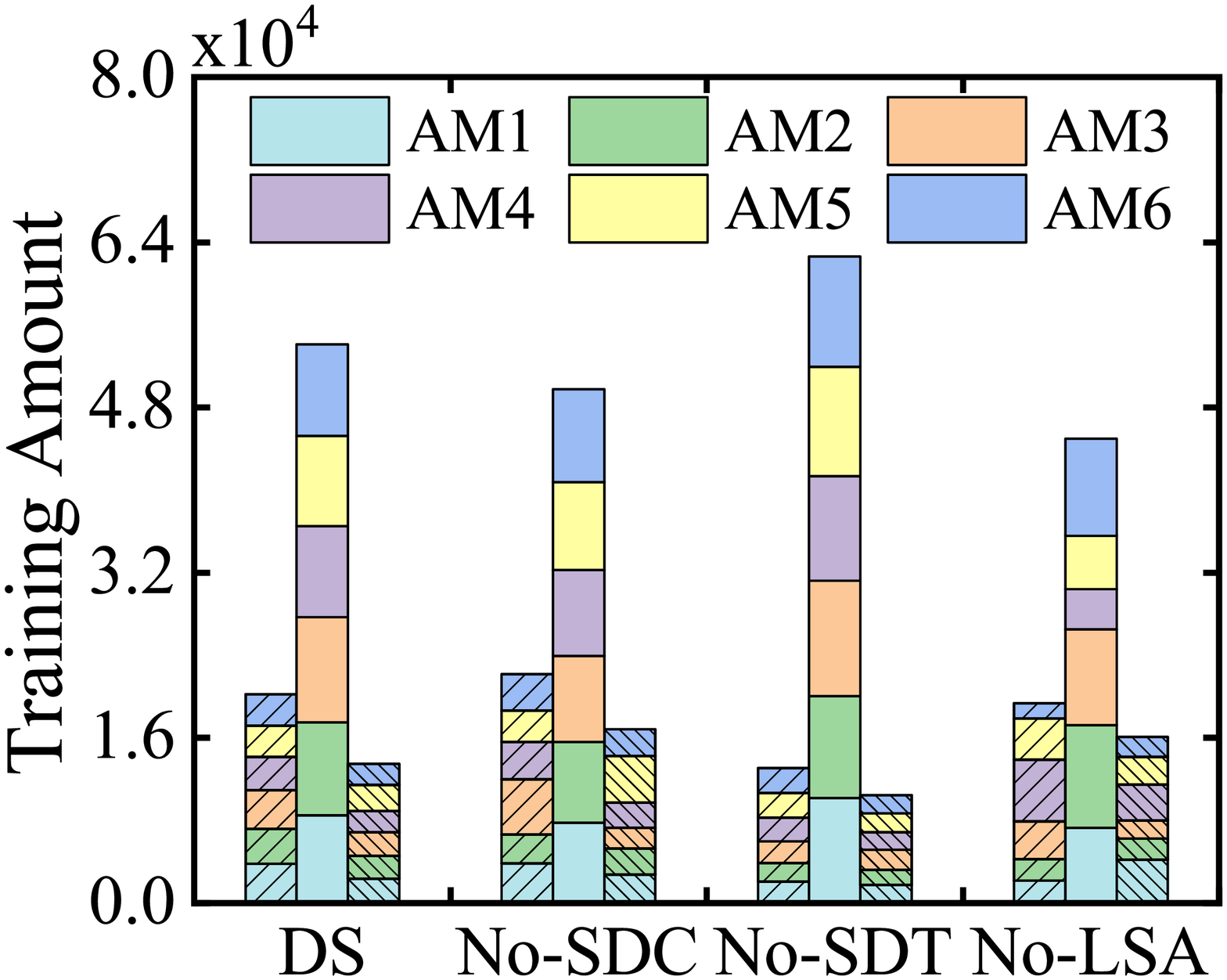}\caption{Data training amount.}
\end{minipage}%
\begin{minipage}[t]{0.33\linewidth}
\centering
\includegraphics[height=4.2cm]{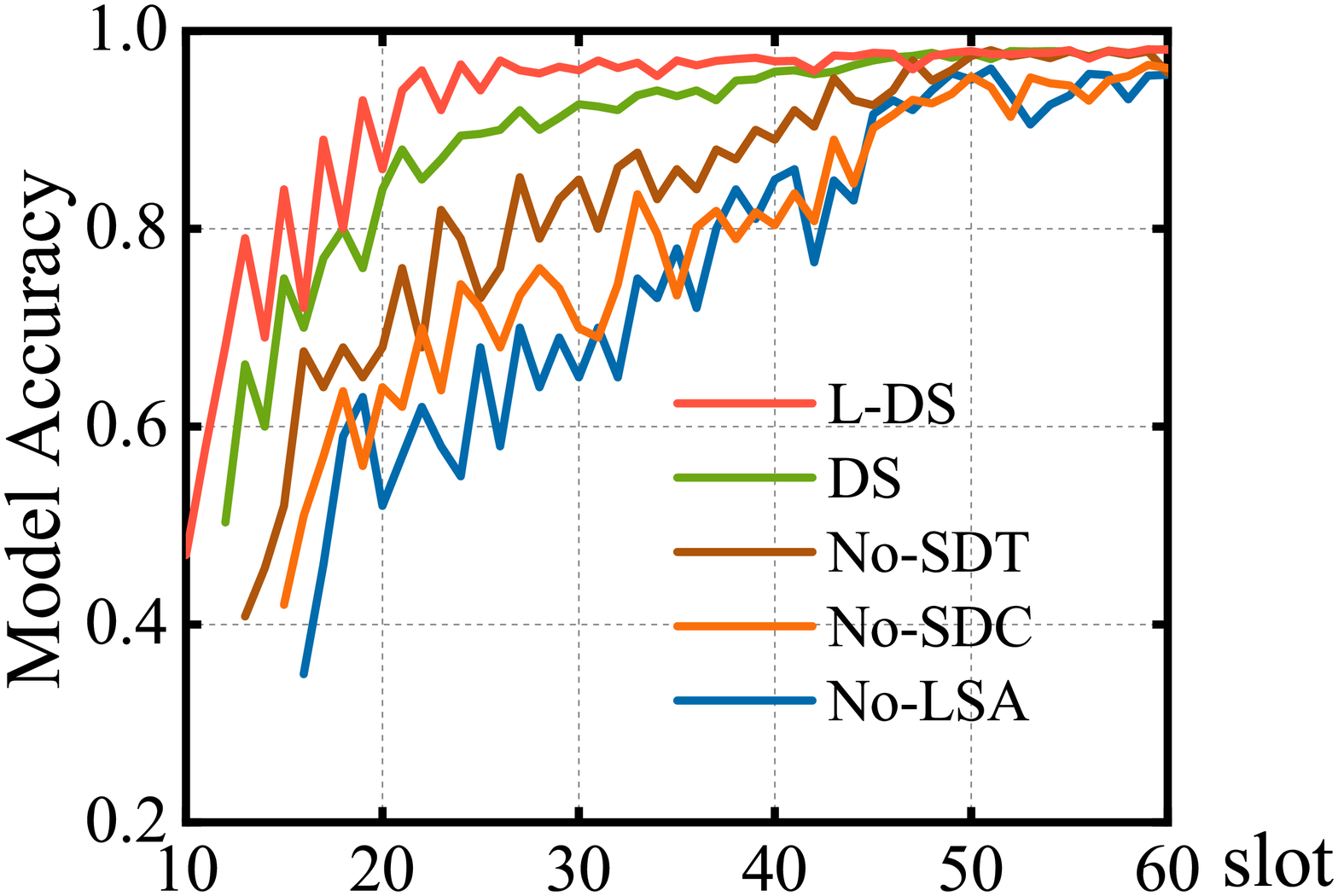}\caption{Model accuracy.}
\end{minipage}%
\label{fig:modularity}
\end{figure*}

To answer this question, we will compare the \emph{DataSche} algorithm (DS) with the following three algorithms:
\begin{itemize}
  \item \emph{NO-SDC} refers to DS without skew-aware data collection (i.e., solving $\mathcal{L}_1$+$\mathcal{L}'_2$)
  \item \emph{NO-SDT} refers to DS without skew-aware data training (i.e., solving $\mathcal{L}'_1$+$\mathcal{L}_2$)
  \item \emph{NO-LSA} refers to DS without long-term skew amendment
\end{itemize}

The evaluation metric is the Standard Deviation (STDEV) and the value set is made of the number of data samples uploaded by six AMs (the number of data samples trained from six AM queues in each MC). Intuitively, a smaller STDEV indicates a more balanced data collection (data training).

Fig. 6 shows the data uploading amount of each AM under four algorithms. Fig. 7 shows the data training amount of each MC under four algorithms (i.e., the three columns of each algorithm respectively refer to the result of MC1, MC2 and MC3, and each column is made of the trained data amount from six AMs).
From Fig. 6, we can easily observe that NO-SDC, NO-SDT and NO-LSA presents a highly imbalanced data upload amount among AMs. Their STDEVs of data collection are 1093, 817 and 1021, while ours is only 311.
From Fig. 7, NO-SDC and NO-LSA presents a clearly imbalanced data training amount from different AMs especially in MC1 and MC3. Although NO-SDT achieves a relatively balanced data training in each MC, the total data training amount among MCs under this algorithm differs greatly. The main reason is that MC2 (8 CPU cores) has more computing capacity and could ``borrow" some data samples from MC1 (2 CPU cores) and MC3 (2 CPU cores) to fully utilize its capacity, and consequently the total data training amount in MC2 is the highest. However, this situation could prolong the convergence time of global ML model, since the global ML model parameters are the average of all the local parameters and the local parameters derived by MC1 and MC3 are undesirable (i.e., the total trained data in MC1 and MC3 is insufficient). As a conclusion, the results in Fig. 6 and Fig. 7 not only indicate that the skew-aware data collection, skew-aware data training and long-term skew amendment all takes effects on averting the skewed data training, but also highlights the importance of long-term skew amendment (i.e., the performance of NO-LSA is the worst). In other words, the capacity heterogeneity of MCs indeed leads to a severe data skew issue over time without the long-term skew amendment.

\textbf{(2) Will the skewed data training adversely impact the trained model accuracy?}

To answer this question, we will still compare the \emph{DataSche} algorithm with the above three algorithms, and the evaluation metric is the model accuracy, which is defined as follows.
\begin{align}
\text{Model accuracy} = \frac{\text{The number of ``good" traffic prediction}}{\text{{The number of test samples}}},\notag
\end{align}
where a ``good" traffic prediction refers to the ``normalized prediction error" is less than 15\%. Here, the normalized prediction error is defined as the gap between the predicted and actual traffic record compared with the actual one. The reason why we choose 15\% is that we offline exploit all the training datasets assigned to AMs to train the LSTM network as mentioned in Section V-A, evaluate the offline trained model with the created testing dataset, and derive that the worst prediction precision is 15\%. In other words, given any a test sample, if the prediction precision achieved by an online trained model is larger than 15\%, then it is worse than the worst case of the offline trained model, and hence we consider it as a ``bad" prediction. In this context, according to the definition of model accuracy, if the model accuracy of an online trained model is close to 1, then we can argue that its prediction performance is approximate to that of the offline trained model. In addition, we also exploit the model accuracy to indicate the training convergence time (i.e., the time slot after which the model accuracy of an online trained model is smooth and steady).

Fig. 8 shows the accuracy of online trained model under different algorithms, where the x-axis refers to the time slot. For example, the x-axis value is 30 indicating the model has been trained for 30 time slots. We can find that with the training time increasing, the model accuracy under each algorithm is also increasing, and all of them will achieve a high accuracy in the end. In other words, if the number of trained data is sufficiently large (i.e., a long training time), the skewed data training would not impact the trained model accuracy greatly. Despite of it, we can also find that our DS algorithm achieves a rapid growth in the beginning (i.e., slot 15 to 25) and a highly steady accuracy (i.e., the performance fluctuation is within 5\%) after slot 30. To achieve the similar performance, NO-SDC, NO-SDT and NO-LSA all roughly requires 50 time slots. In this context, we can conclude that the skewed data training will prolong the training convergence time. Since our proposed algorithm can effectively alleviate it from both short-term (i.e., skew-aware data collection and training per slot) and long-term (i.e., long-term skew amendment) perspective, it can achieve a better performance.

\textbf{(3) Will the \emph{Learning-aid DataSche} algorithm be better than the \emph{DataSche} algorithm?}
\begin{figure*}[tt]
\centering
{\subfigcapskip = -0.1cm
\subfigure[\small Framework cost]{\centering
\includegraphics[height=3.4cm]{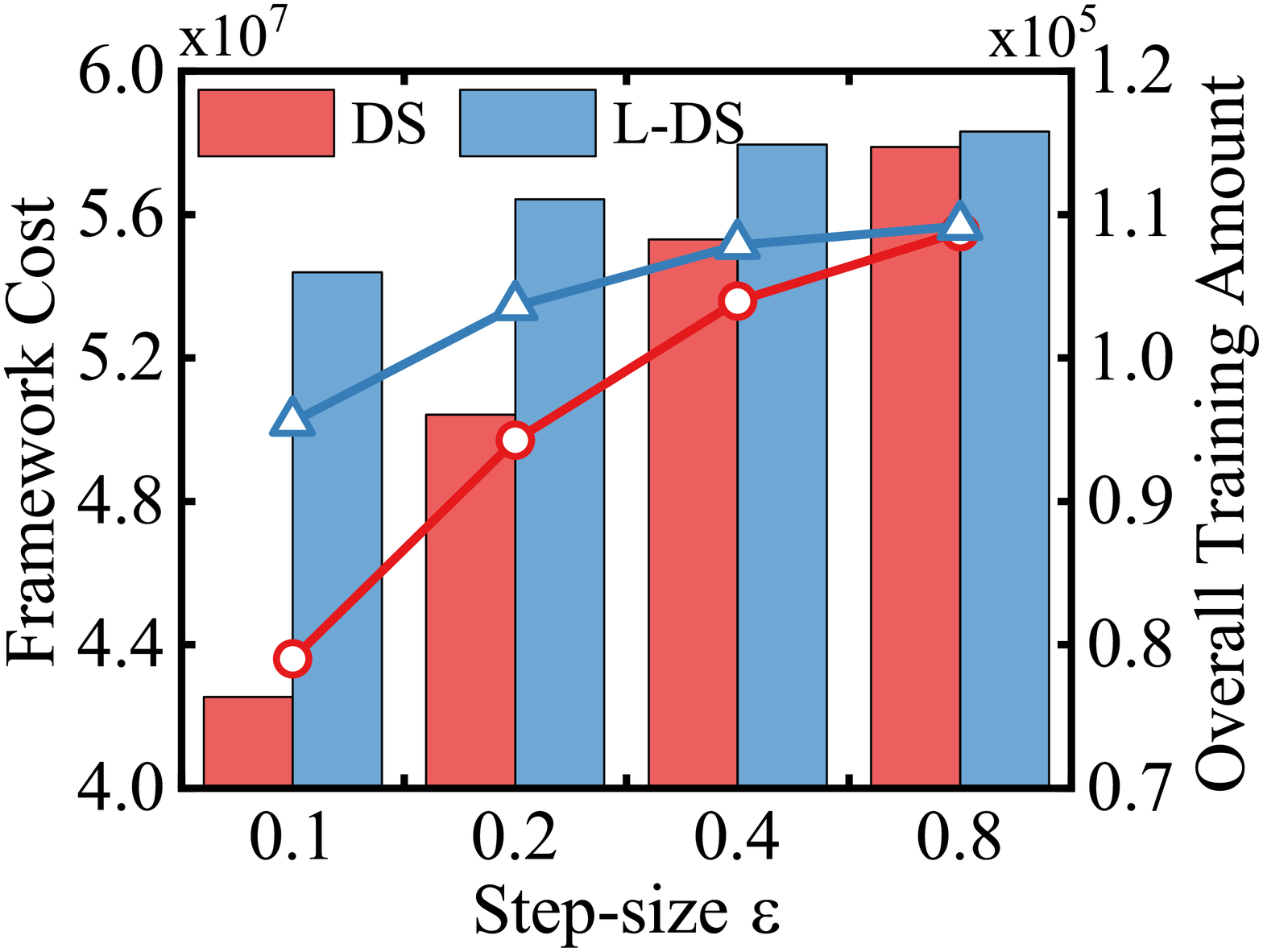}}
\subfigure[\small AM queue backlog]{\centering
\includegraphics[height=3.3cm]{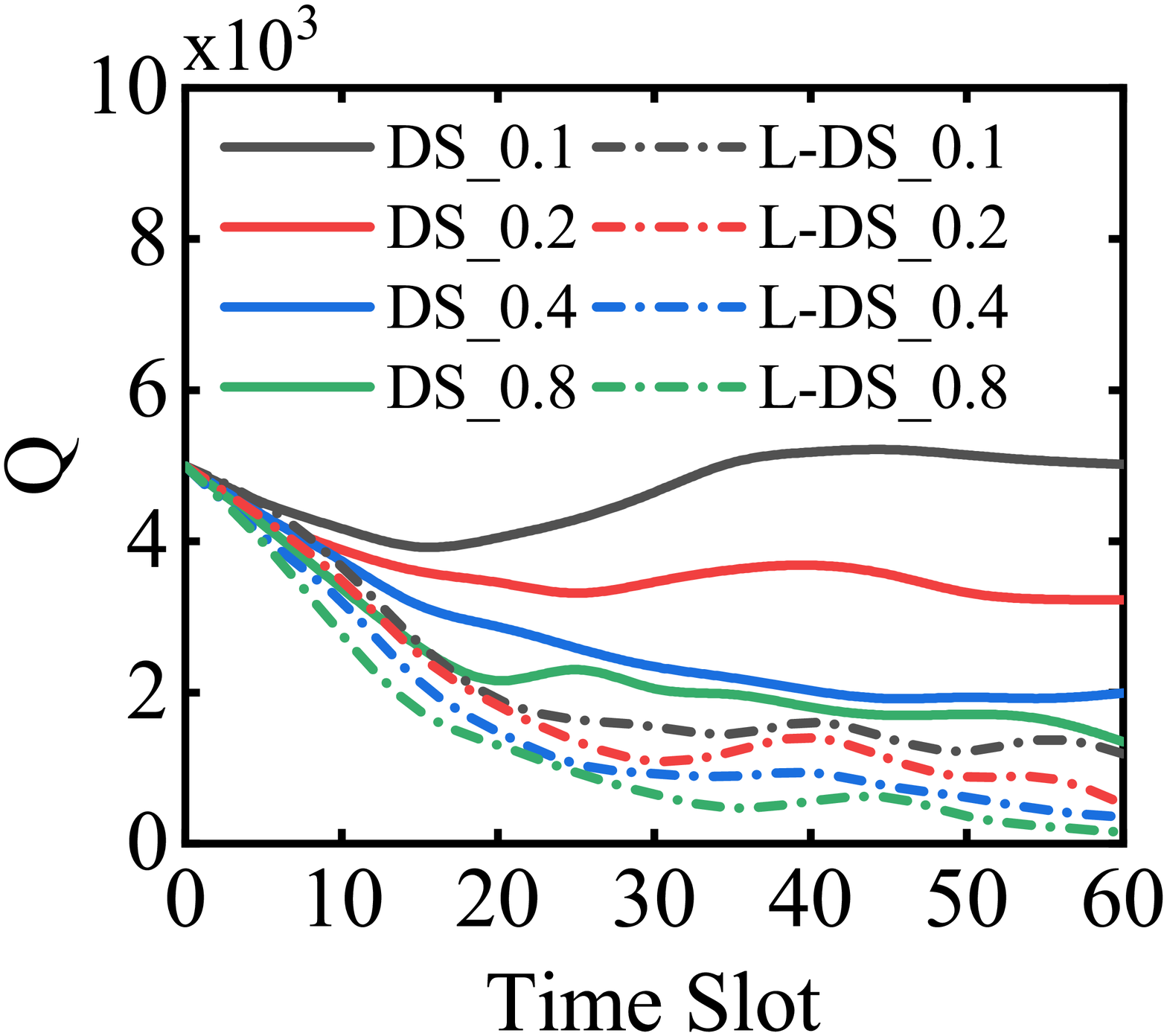}}
\subfigure[\small MC queue backlog]{\centering
\includegraphics[height=3.3cm]{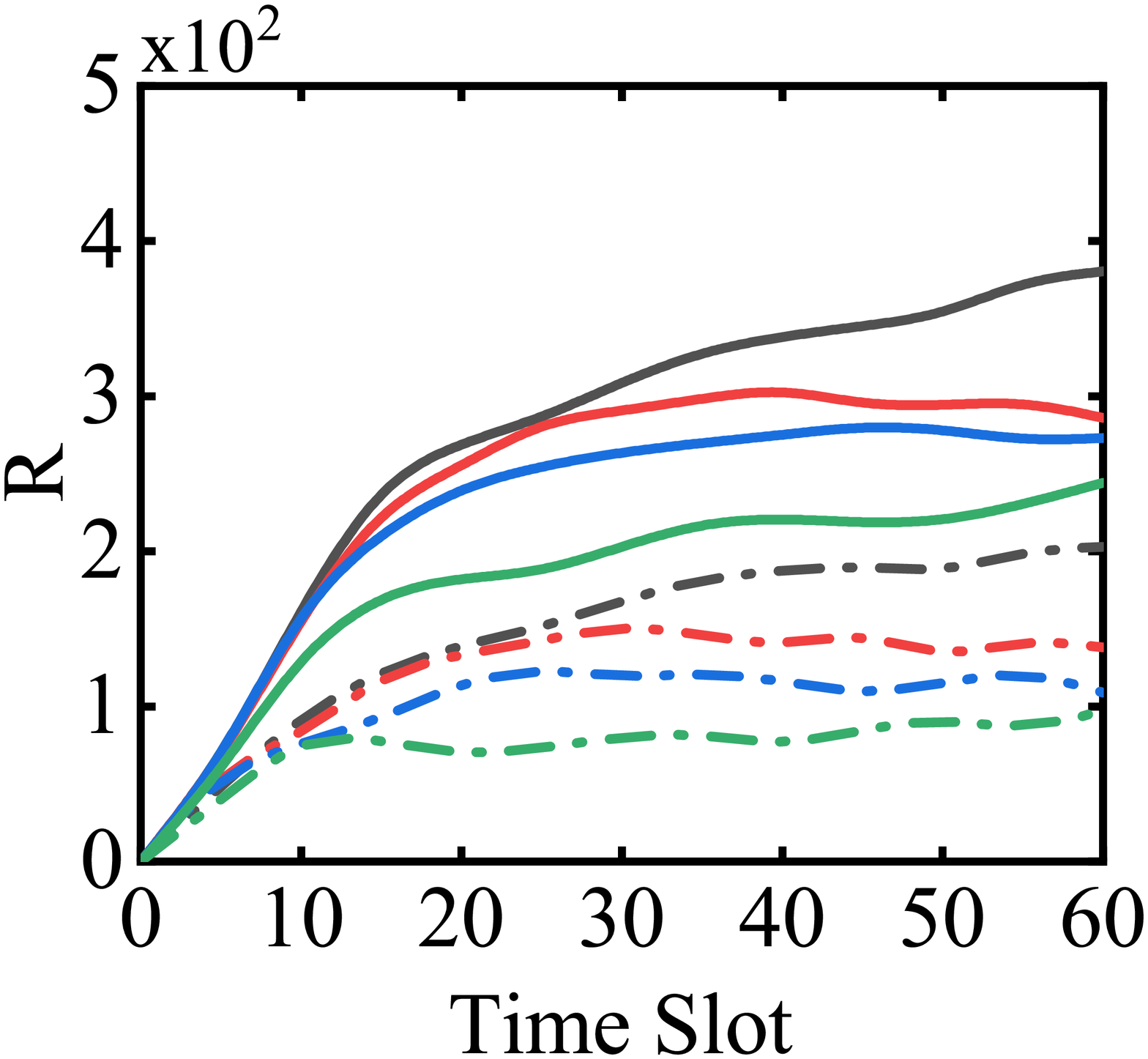}}
\subfigure[\small Degree of skewness]{\centering
\includegraphics[height=3.3cm]{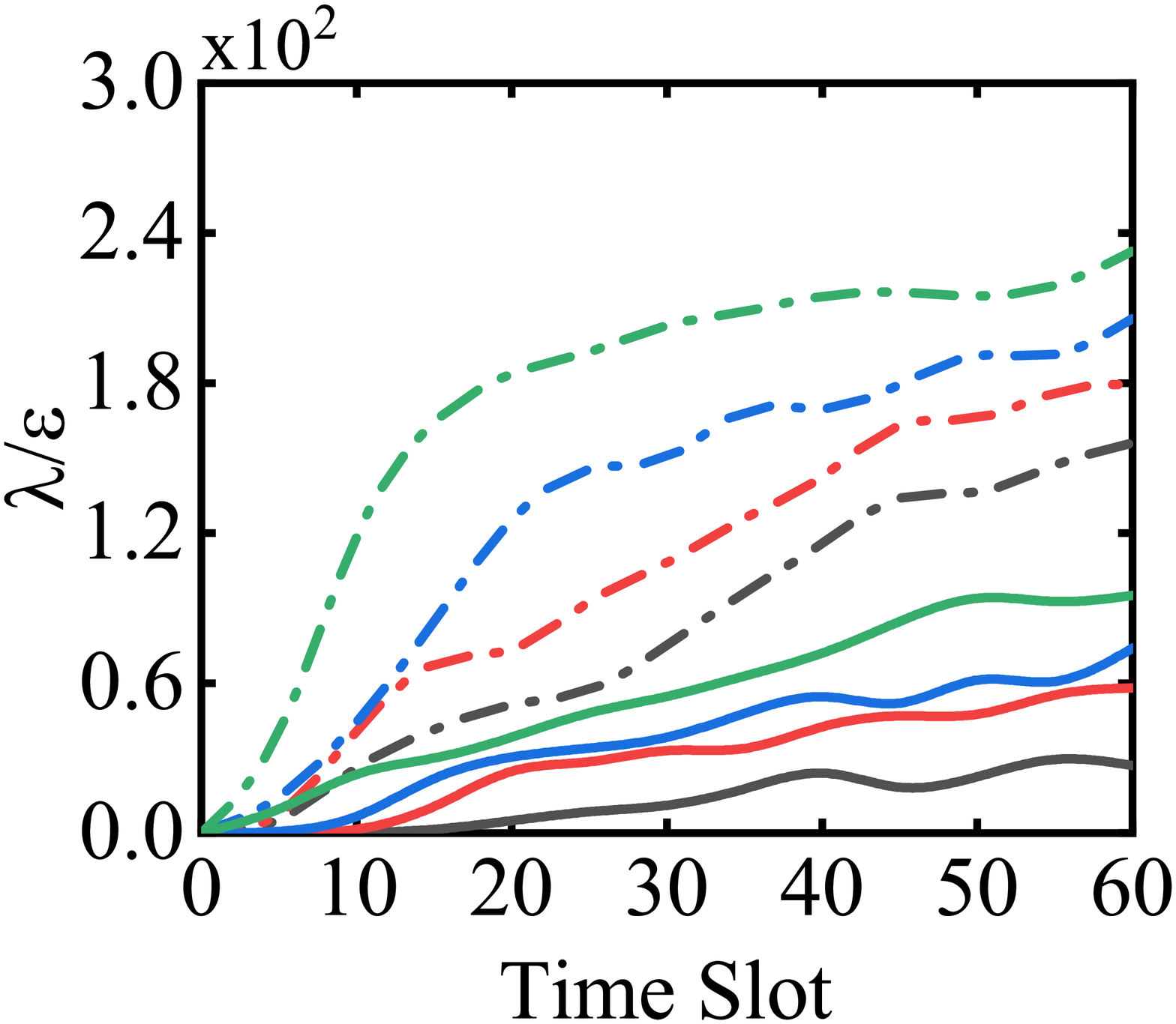}}
}
\caption{The performance comparison between the \emph{DataSche} algorithm (DS) and the \emph{Learning-aid DataSche} algorithm (L-DS); ``0.1, 0.2, 0.4 and 0.8" refer to the values of step-size $\varepsilon$.}
\label{fig:results}
\end{figure*}

\begin{figure*}[tt]
\centering
{\subfigcapskip = -0.1cm
\begin{minipage}[t]{0.33\linewidth}
\centering
\subfigure[\small Data training amount]{\centering
\includegraphics[height=4.2cm]{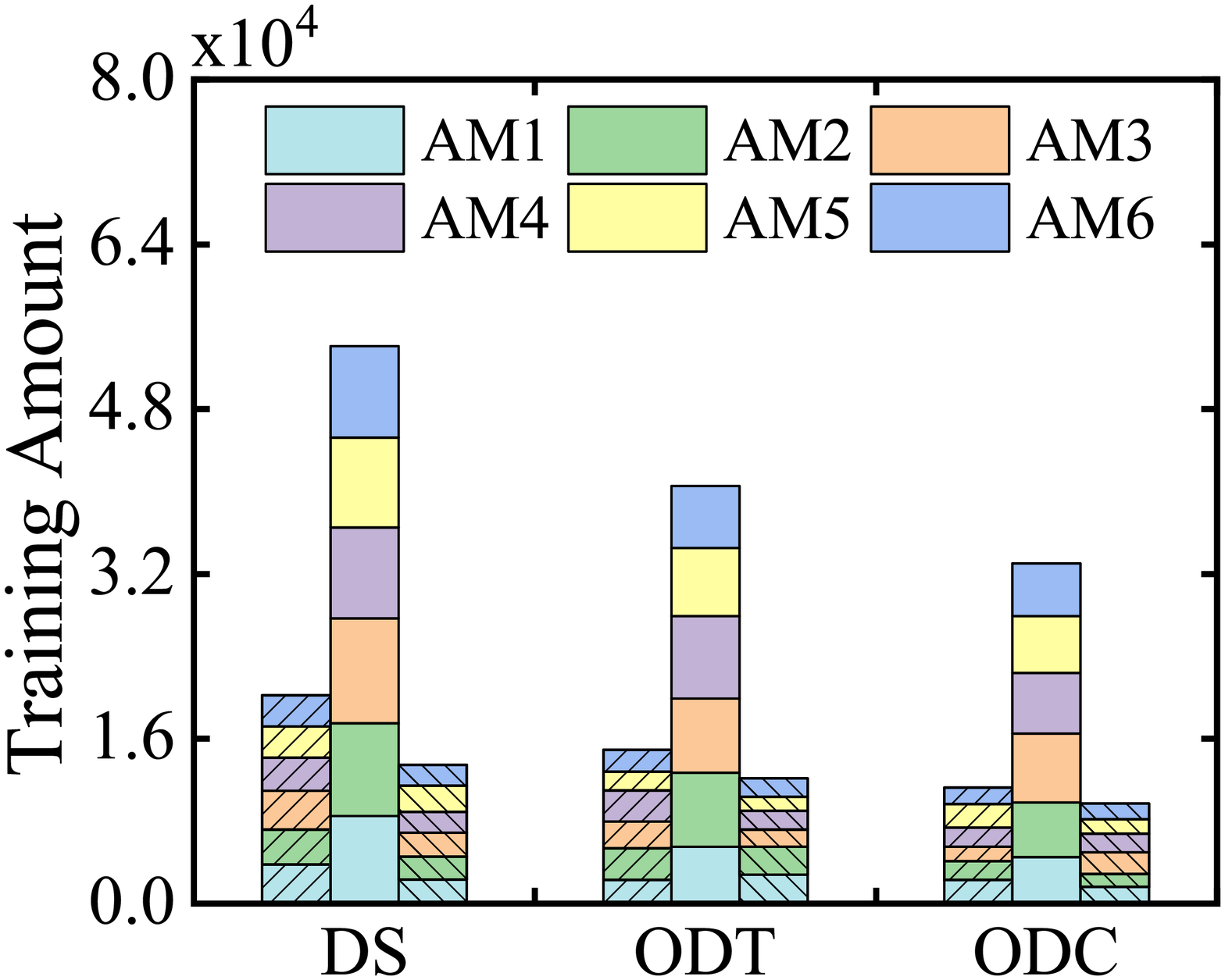}}
\end{minipage}%
\begin{minipage}[t]{0.33\linewidth}
\centering
\subfigure[\small Unit training cost]{\centering
\includegraphics[height=4.2cm]{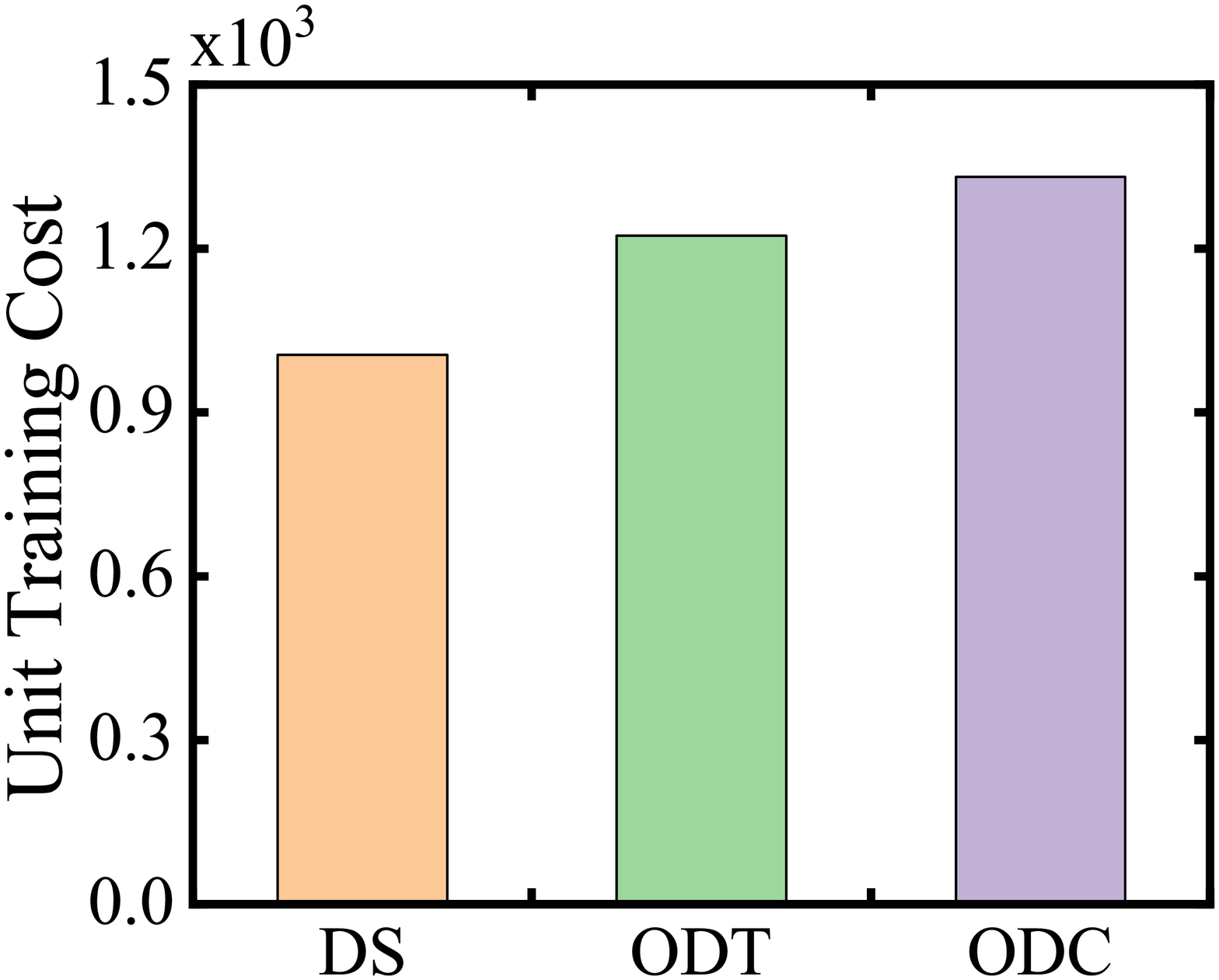}}
\end{minipage}%
\begin{minipage}[t]{0.33\linewidth}
\centering
\subfigure[\small Model accuracy]{\centering
\includegraphics[height=4.25cm]{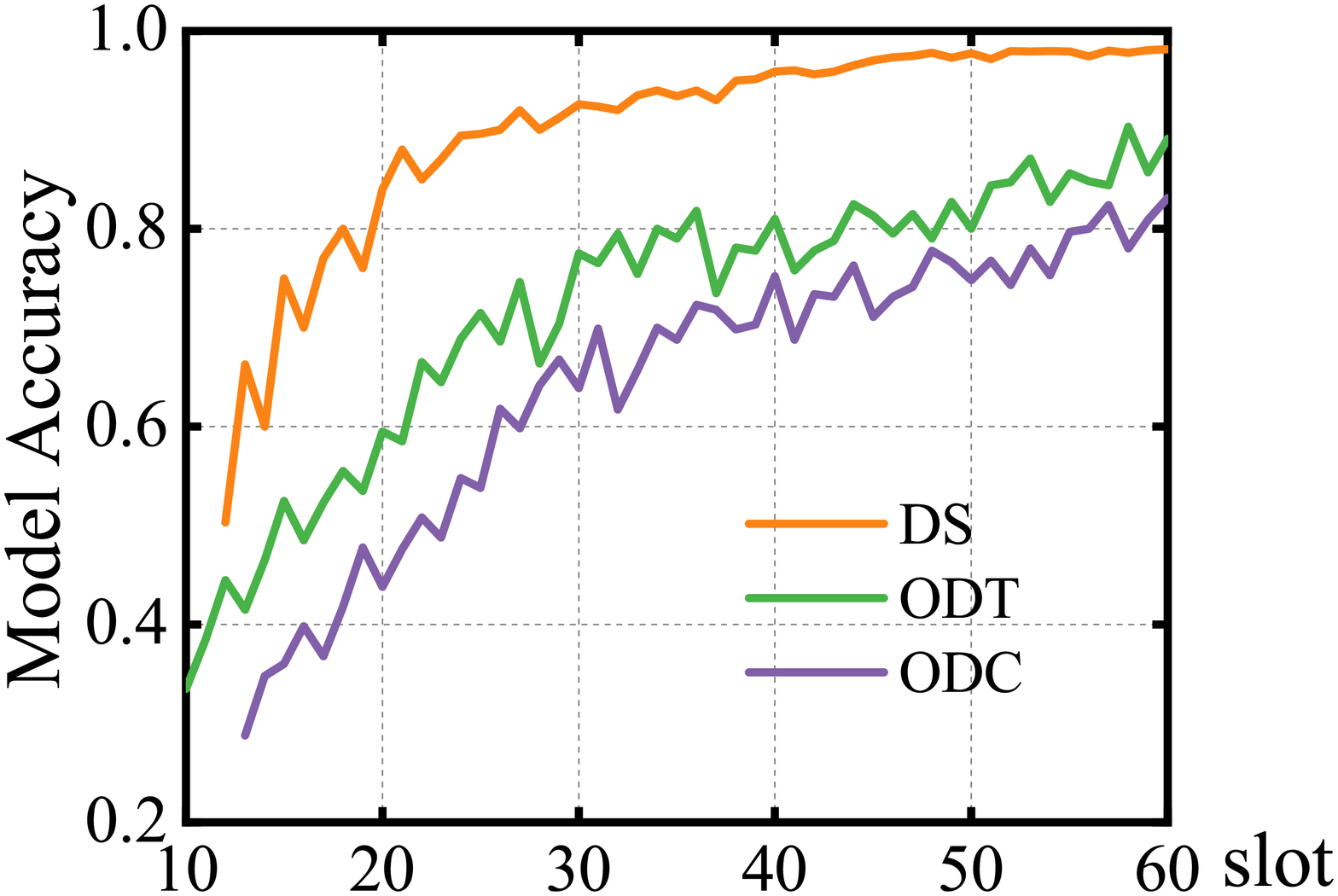}}
\end{minipage}%
}
\caption{The performance comparison among different algorithms.}
\label{fig:comparison}
\end{figure*}

To answer this question, we will compare the \emph{DataSche} algorithm with the \emph{Learning-aid DataSche} algorithm under different step-size $\varepsilon$, in terms of the framework cost, total queue backlog of AMs, total queue backlog of MCs and the degree of data skewness.

The evaluation results\footnote{We only depict the result of virtual queue $\lambda/\varepsilon$ standing for constraint (11) (i.e., the larger backlog refers to the more trained samples compared with the expected value), and that for constraint (10) shares the similar trend.} are given in Fig. 9. We can see that with the value of step-size $\varepsilon$ increasing, the framework costs of both algorithms are increasing and the total queue backlogs of both AMs and MCs are decreasing. These phenomena are in accordance with our theoretical analysis. From Fig. 9(a), the framework cost (the column) of L-DS is larger than that of DS especially when $\varepsilon$ is small (i.e., the cost of DS is roughly 75\% of that of L-DS). The main reason is that L-DS can effectively cope with the queue backlog under a small step-size, which can be confirmed by the value and trend of overall data training amount (the line) in Fig. 9(a) as well as the total queue backlog of both AMs and MCs in Fig. 9(b) and Fig. 9(c). For example, the number of data trained by L-DS will be 1.2x compared with that by DS when $\varepsilon=0.1$. In addition, we can see that L-DS and DS will train the similar number of data samples when $\varepsilon=0.8$. In other words, a large $\varepsilon$ value actually pushes forward the data collection and training, which can be confirmed by the total queue backlog of both AMs and MCs in Fig. 9(b) and Fig. 9(c). For example, when $\varepsilon=0.8$ the total queue backlog of AMs and the total queue backlog of MCs under each algorithm are both the least (i.e., more data collection and training).

From Fig. \ref{fig:results}(d), we can find that the degree of data skewness of both algorithms given any step-size $\varepsilon$ will be bounded, and this phenomenon indicates that the proposed two algorithms can effectively avert the skewed data training. In addition, we can see that L-DS involves more skewed data training compared with DS (i.e., a large accumulated value of $\lambda/\varepsilon$). This is because the empirical Lagrange multipliers in L-DS serve as ``virtual" queue backlog, which gives more weight on data collection and training rather than skew amendment. Therefore, collecting and training more data in each EC may easily lead to a data skew issue over time. Despite of it, as for the accuracy of online trained model, L-DS can achieve 1.5x fast convergence time and a higher steady performance compared with DS. This phenomenon once more indicate that training sufficiently large number of data can actually counteract the negative effect of the data skew issue. To sum up, we can conclude that the performance of L-DS is preferable in practice, especially when the step-size $\varepsilon$ is small (i.e., an acceptable framework cost and a good trained model with a low convergence time).

\begin{figure*}[tt]
\centering
{\subfigcapskip = -0.1cm
\subfigure[\small Total training amount]{\centering
\includegraphics[height=3.4cm]{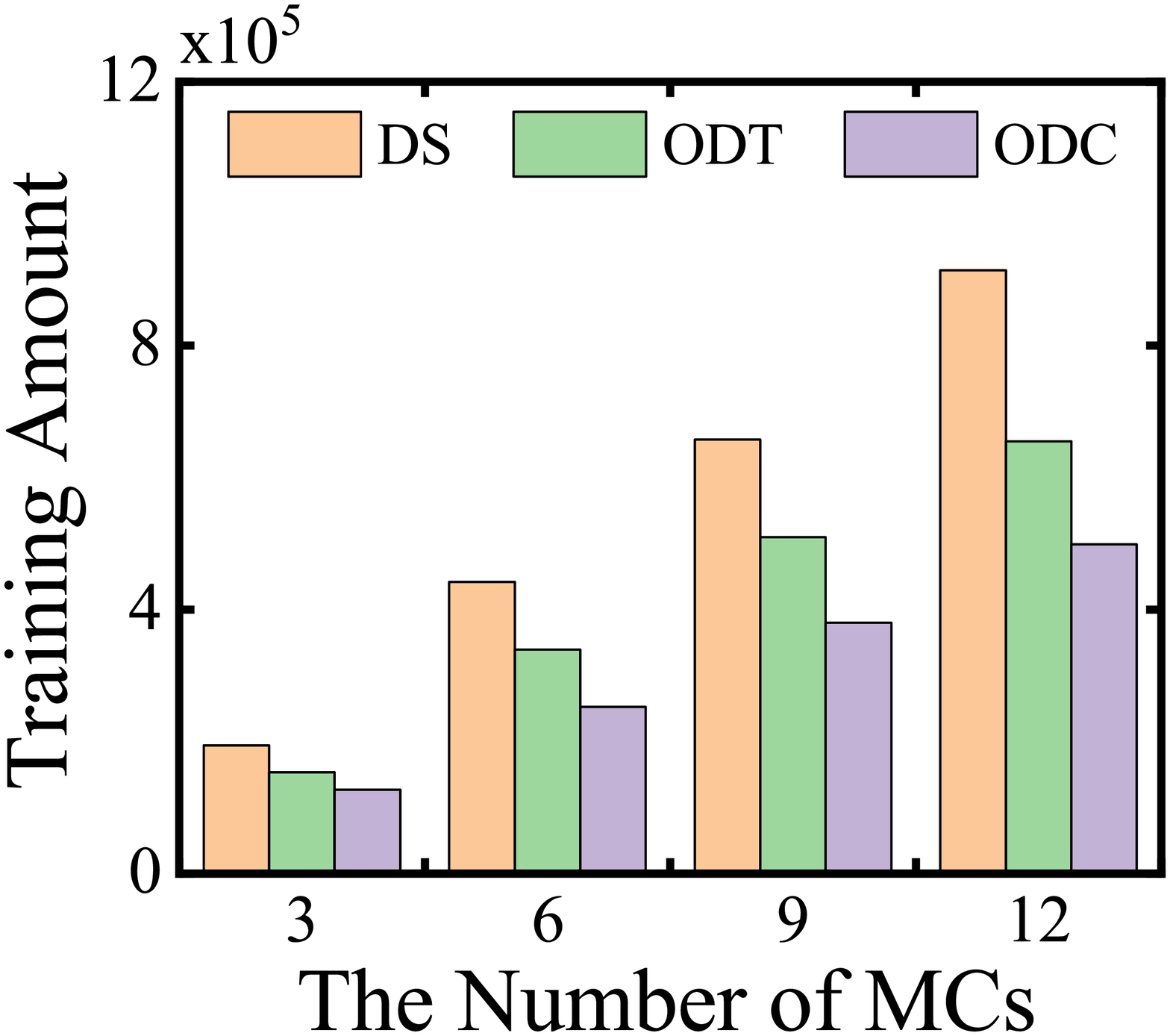}}
\subfigure[\small Unit training cost]{\centering
\includegraphics[height=3.4cm]{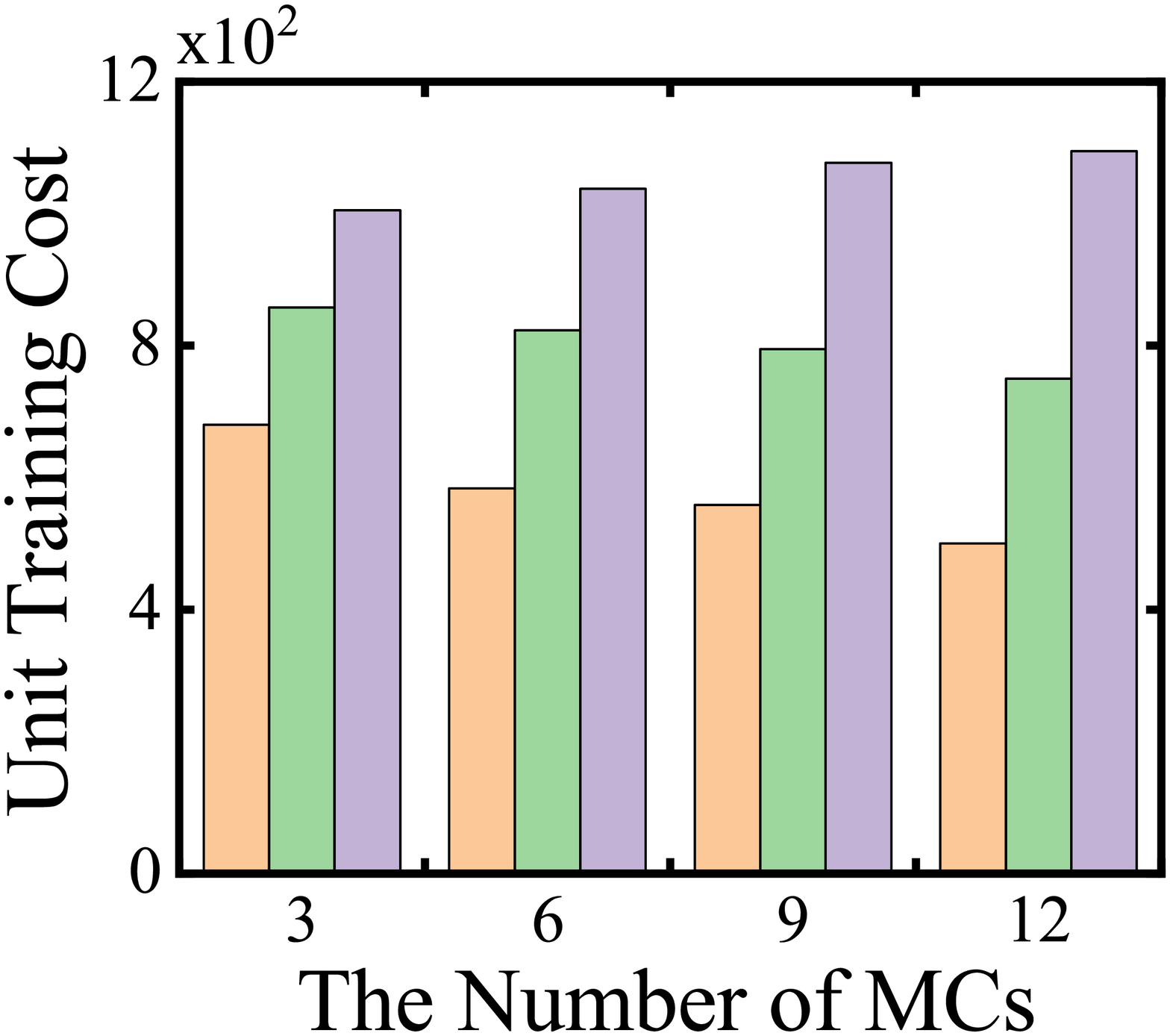}}
\subfigure[\small MSE (MC=3)]{\centering
\includegraphics[height=3.4cm]{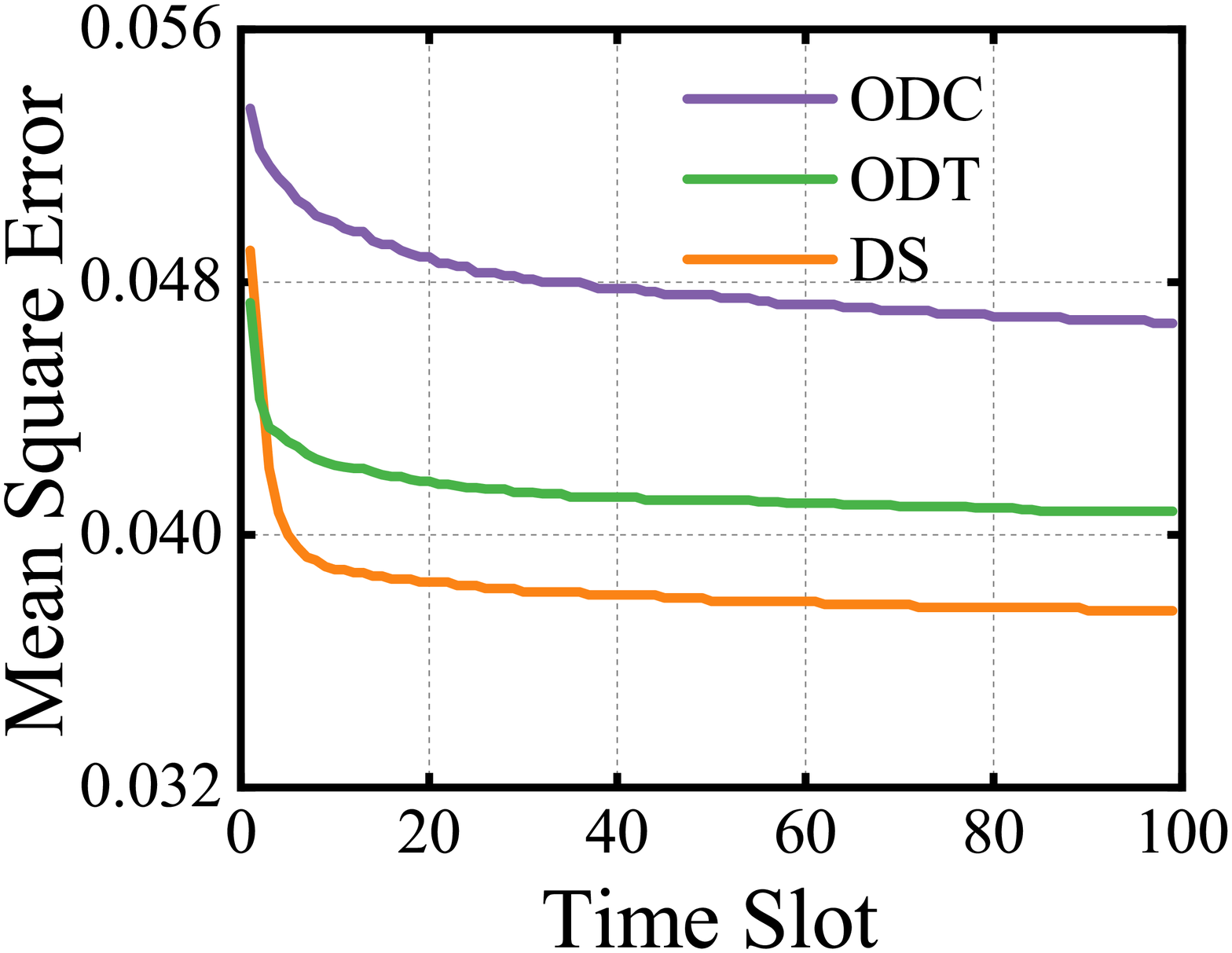}}
\subfigure[\small MSE (MC=12)]{\centering
\includegraphics[height=3.4cm]{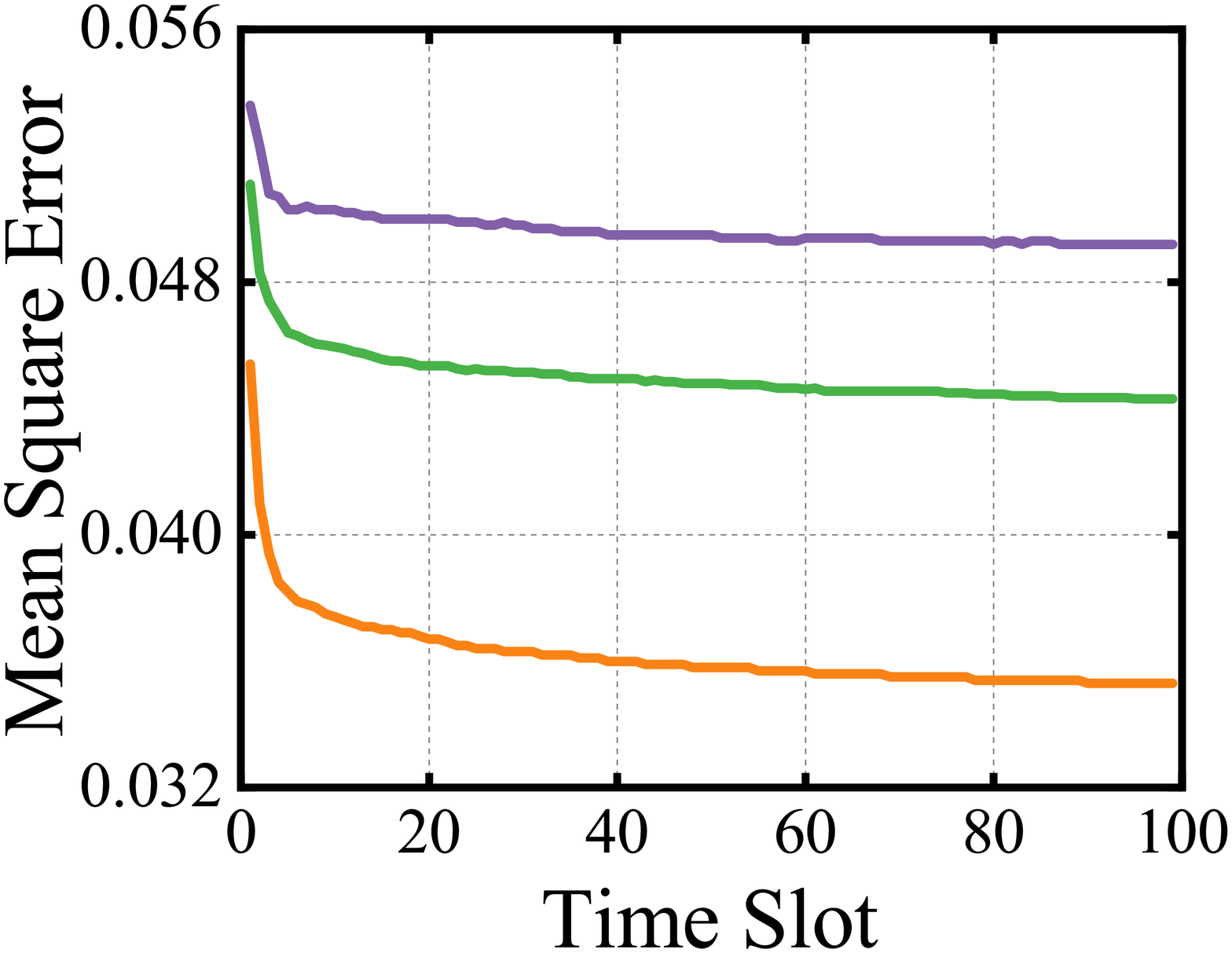}}
}
\caption{The performance comparison among different algorithms with different numbers of MCs (AM=20).}
\label{fig:results2}
\end{figure*}

\begin{figure*}[tt]
\centering
{\subfigcapskip = -0.1cm
\subfigure[\small Total training amount]{\centering
\includegraphics[height=3.4cm]{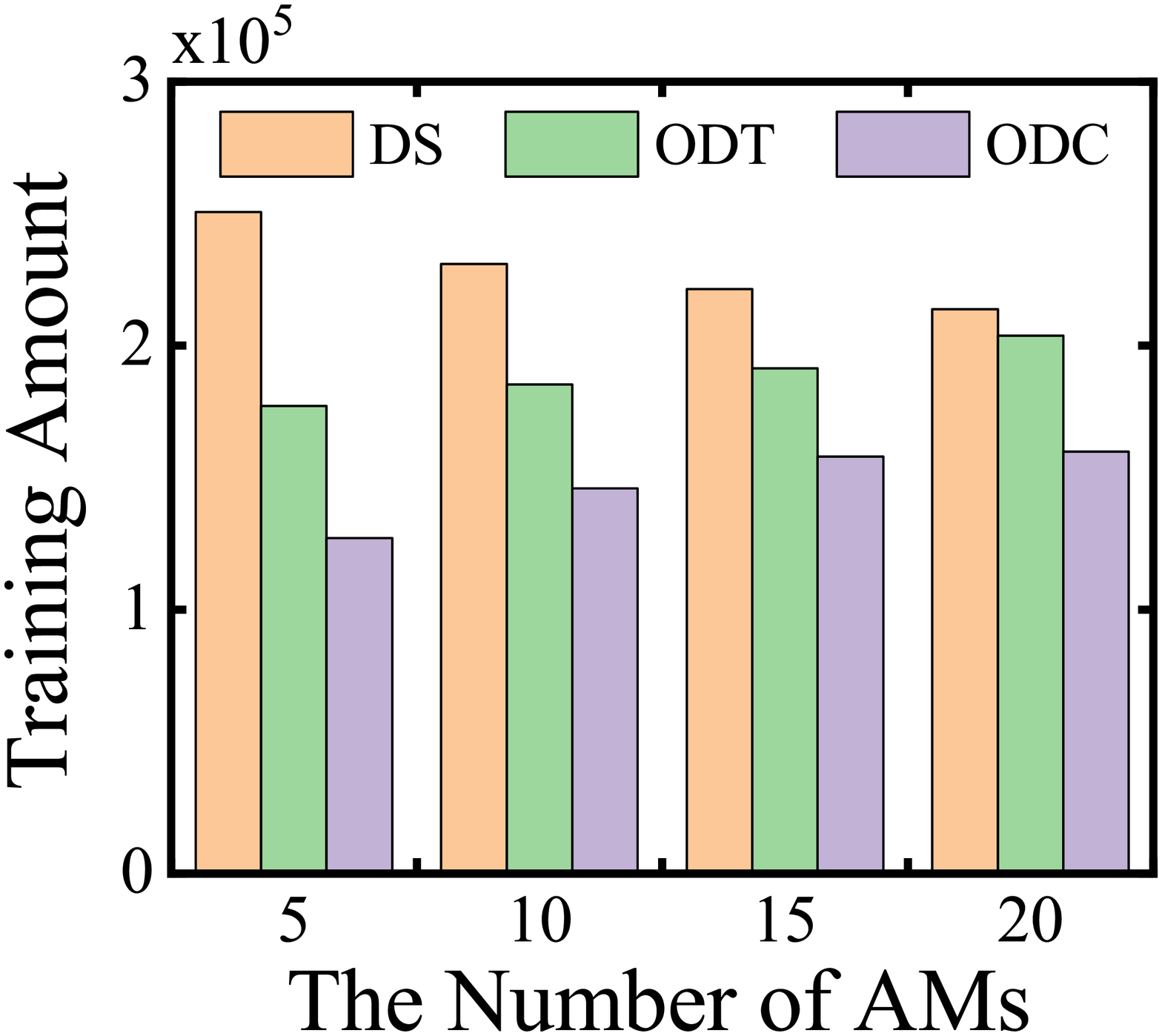}}
\subfigure[\small Unit training cost]{\centering
\includegraphics[height=3.4cm]{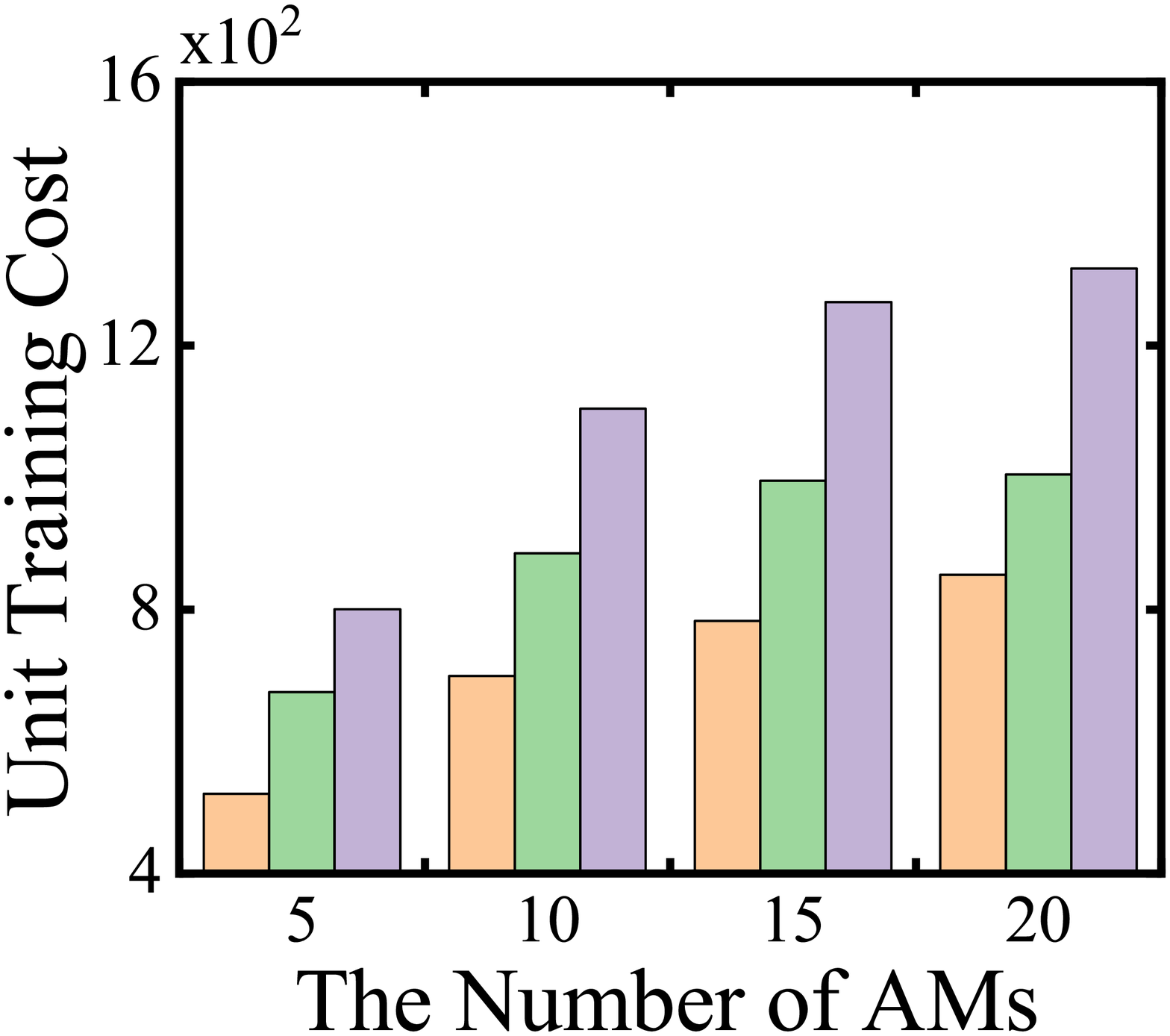}}
\subfigure[\small MSE (AM=5)]{\centering
\includegraphics[height=3.4cm]{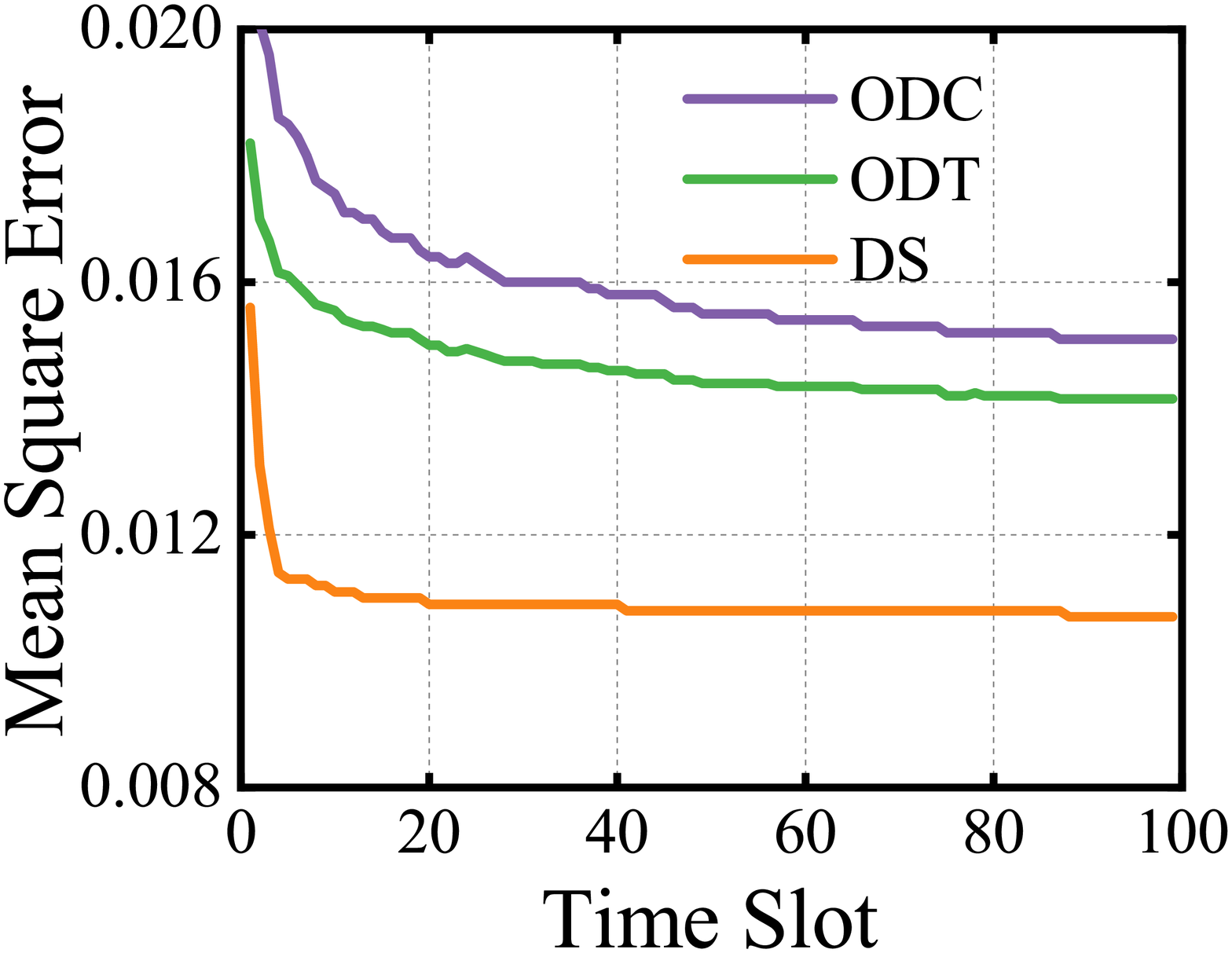}}
\subfigure[\small MSE (AM=20)]{\centering
\includegraphics[height=3.4cm]{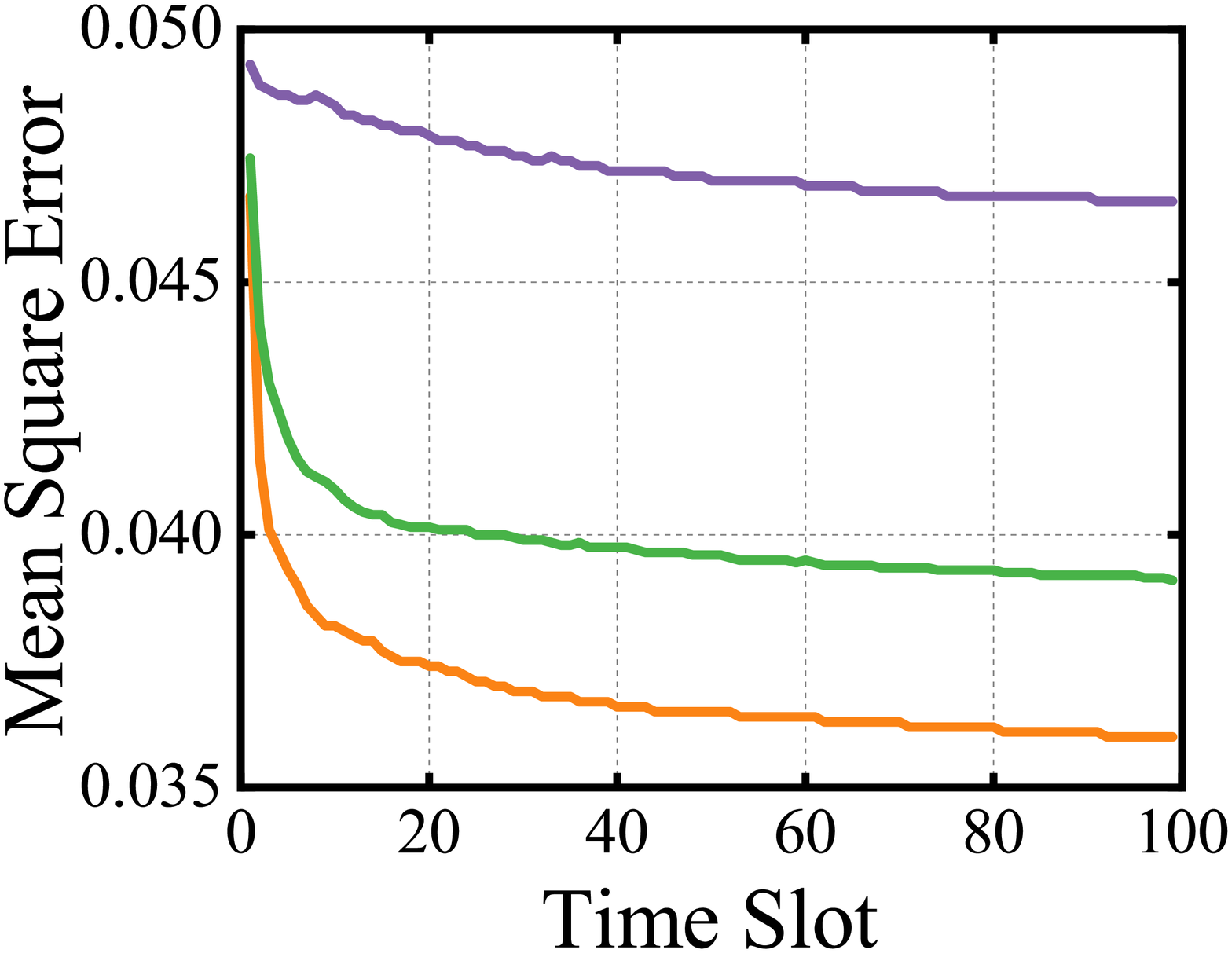}}
}
\caption{The performance comparison among different algorithms with different numbers of AMs (MC=5).}
\label{fig:results3}
\end{figure*}

\textbf{(4) What is the performance of the \emph{DataSche} algorithm if taking alternative design principle into account such as given AM\! --\! MC connections and worker non-cooperation?}

To answer this question, we will compare the \emph{DataSche} algorithm with the following two algorithms.
\begin{itemize}
  \item \emph{ODT} refers to the algorithm that only involves skew-aware data training (i.e., solving $\mathcal{L}'_2$) and takes fixed data collection into account (i.e., each AM is assigned to a given MC in advance, and each MC will evenly allocate the connection duration to the connected AMs), which embodies the feature of geo-distributed machine learning
  \item \emph{ODC} refers to the algorithm that only involves skew-aware data collection (i.e., solving $\mathcal{L}'_1$) and does not consider worker cooperation (i.e., no data offloading), which embodies the feature of federated learning
\end{itemize}

The evaluation metric is the total training amount, unit training cost (i.e., the framework cost divided by the total training amount) and model accuracy.

The evaluation results are given in Fig. 10. Fig. 10(a) similar to Fig. 7 presents the data training amount of each MC under three algorithms, and we can find that the training amount of each MC under DS is more than that of ODT and ODC. Specifically, DS respectively achieves 29.8\% and 62.6\% more trained data in total compared with ODT and ODC. This result particularly highlights the importance of worker cooperation (i.e., the performance of ODC is the worst). In addition, similar to DS, ODT and ODC also presents a relatively balanced data training amount from different AMs, which further indicates that both the skew-aware data collection and the skew-aware data training take effects on averting the skewed data training.

Fig. 10(b) presents the unit training cost of each algorithm, and DS respectively achieves 21.7\% and 32.4\% reduction compared with ODT and ODC. The reasons are as follows. ODT requires more data transmission between MCs to achieve the skew-aware data training, which leads to more data offloading cost. ODC fails to fully utilize the computing capacity of MCs (i.e., worker non-cooperation), which leads to a large number of remaining data uploaded from AMs (i.e., a large queue backlog). In other words, it trains a limited number of data with a lot of data collection cost.

Fig. 10(c) presents the accuracy of online trained model under three algorithms, and we can derive that DS respectively achieves 10.2\% and 18.1\% performance gain compared with ODT and ODC after slot 40. The main reason is that DS will train more data from each AM as shown in Fig. 10(a), and therefore its online trained model could better capture the features of the dataset assigned to each AM and achieve a good prediction precision. In summary, we can conclude our proposed online algorithm with adaptive data collection and worker cooperation can achieve better performance.

\subsection{Simulation Evaluation}\label{simulation}
We also conduct large-scale simulations to evaluate the scalability of the proposed \emph{DataSche} algorithm. To this end, we exploit the Opportunistic Network Environment (ONE)
simulator \cite{ONE} to create a simulation scenario. Specifically, we consider a 1km$\times$1km square area and create two groups of nodes to respectively indicate AMs and MCs. Each node is randomly deployed in the area, so as to simulate the network topology. We adopt the similar setting principle as that for testbed with different baseline values: the baseline of MC computing capacity is randomly selected from \{2, 5, 10\} CPU cores@3.0GHz with dynamics following Fig. 5(c); that of the unit cost $c_{ij}(t)=400$, $e_{jk}(t)=60$, $p_j(t)=100, \forall i, \forall j, \forall k, \forall t$; that of AM\! --\! MC transmission capacity is randomly selected from \{0.5, 1.5\} Mbps; that of MC\! --\! MC transmission capacity is 3 Mbps. 

The MCs will cooperatively train an ML model (i.e., a deep neural network: six full connection layers with neuron 512, 256, 128, 128, 5 and 5) for base station power allocation with the synthetic dataset \cite{sun2018learning} (i.e., network control function). We consider 20 power allocation scenarios with different number of users, and generate 100000 data samples for each scenario. We exclusively assign one of the 20 scenarios to an AM in the simulation. We adopt the similar method to calculate $\rho=1.9\times10^{7}$ cycles. Each time slot is set to 1 second and the number of slots is 100 in the simulation.

To evaluate the scalability of our proposed algorithm, we also compare it with the alternative solution ODT and ODC, in terms of different numbers of AMs and MCs. The evaluation metric is the total training amount, unit training cost, mean square error (MSE) of the online trained model (i.e., the value of the loss function) and algorithm running time. Note that the MSE is derived as follows. We first run the simulation to derive the number of trained data from each AM in each MC per slot, which is regarded as the data training records, and then build a distributed learning program to train the DNN model for power allocation in terms of the derived records.

The evaluation results are given in Fig. \ref{fig:results2}, Fig. \ref{fig:results3}, Table II and Table III. As we can see from Fig. \ref{fig:results2}(a) and Fig. \ref{fig:results2}(b), given the number of AM, the total training amount of each algorithm achieves a clear increase and the unit training cost of DS and ODT slightly decreases as the number of MCs increases. The reason is as follows. The more number of MCs indicates the more number of training workers, which intuitively leads to the more trained data. However, it also indicates the more chance to conduct data offloading between MCs, which to some extent reduces the data transmission between AM and MC, and accordingly achieves a smaller unit training cost, since the unit transmission cost between MCs is much smaller than that between MC and AM. Numerically, DS respectively achieves on average 41.1\% and 76.4\% cost reduction compared with ODT and ODC. This is because that our algorithm makes full use of the computing capacity of each MC compared with ODC, and it will not introduce too much data offloading between MCs to facilitate long-term skew amendment.
In this context, the MSE achieved by DS is clearly better than the others as shown in Fig. \ref{fig:results2}(c) and Fig. \ref{fig:results2}(d). For example, when MC=3, the MSE achieved by DS reveals a rapid decrease and a steadily small value (i.e., 0.037) in the end, which is 8.24\% and 24.2\% better than that of ODT and ODC, respectively.

As we can see from Fig. \ref{fig:results3}(a) and Fig. \ref{fig:results3}(b), given the number of MC, the total training amount of DS decreases and the unit training cost increases with the increasing number of AM. This is because DS takes the skew-aware training from both short-term and long-term perspective into account, which makes the number of trained data samples smaller while making the trained data samples from different AMs more balanced. Indeed, as we can see in Fig. \ref{fig:results3}(c) and Fig. \ref{fig:results3}(d), the MSE achieved by DS is much better than the others.

We numerically evaluate the running time of the \emph{DataSche} algorithm with a large number of AMs and MCs. Table II and Table III presents the results by using an Intel Cores i7 Processor@3.0GHz. Overall, we can find that the running time of ODT is the smallest, which is in accordance with the analysis of algorithm complexity, since the complexity of the skew-aware data collection dominates the total running time and ODT does not involve it. In addition, we can find that with the number of AMs or MCs increasing, our running time will increase linearly (e.g., the slope is respectively close to 1 and 2), which indicates that the \emph{DataSche} algorithm could scale well for a large number of AMs and MCs in practice.
\begin{table}[tt]
\centering
\begin{tabular}{|c|c|c|c|c|c|}
\hline
{MC$\equiv$10}    & {AM=20}      & {40}      & {60}     & {80}     & {100}     \\ \hline
{DS}  &  & {0.332}  & {0.428} & {0.559} & {0.685} \\ \hline
{ODT} & {0.026} & {0.027} & {0.025} &
{0.024} & {0.023} \\ \hline
{ODC} & {0.122}  & {0.251}  & {0.343} & {0.462} &  {0.575} \\ \hline
\end{tabular}
\caption{Algorithm running time with different numbers of AMs.}
\end{table}

\begin{table}[tt]
\centering
\begin{tabular}{|c|c|c|c|c|c|}
\hline
{AM$\equiv$100}    & {MC=10}      & {20}      & {30}     & {40}     & {50}     \\ \hline
{DS}  & {0.685} & {1.258}  & {3.206} & {5.238} & {6.934} \\ \hline
{ODT} & {0.023} & {0.052} & {0.115} & {0.354} &  {0.713} \\ \hline
{ODC} & {0.575}  & {1.076}  & {2.351} & {4.894} & {6.277} \\ \hline
\end{tabular}
\caption{Algorithm running time with different numbers of MCs.}
\end{table}

\section{Conclusion}
In this paper, we propose \emph{Cocktail}, an incremental learning framework within a reference 5G network architecture. We build the framework model and formulate an online data scheduling problem to optimize the framework cost while alleviating the data skew issue from the long-term perspective. We exploit the stochastic gradient descent to devise an online asymptotically optimal algorithm, including two optimal policies based on novel graph constructions for skew-aware data collection and data training per time slot. We also improve the proposed algorithm with online learning to speedup the training convergence. Small-scale testbed and large-scale simulations validate its superior performance.
\bibliographystyle{ieeetr}
\bibliography{JSAC2021_Final}

\end{document}